\documentclass[twocolumn]{emulateapj}

\usepackage{graphicx}
\usepackage{amsmath}
\usepackage{array}
\usepackage{tabularx}

\usepackage{natbib}

\def\be{\begin{equation}}
\def\ee{\end{equation}}
\def\ba{\begin{eqnarray}}
\def\ea{\end{eqnarry}}
\def\bal#1\eal{\begin{align}#1\end{align}}

\newcolumntype{W}{>{\raggedright\arraybackslash}X}
\newcolumntype{Y}{>{\raggedleft\arraybackslash}X}
\newcolumntype{Z}{>{\centering\arraybackslash}X}
\begin{document}
\title{Time-dependent, compositionally driven convection in the oceans of accreting neutron stars}
\author{Zach Medin}
\affil{Los Alamos National Laboratory, Los Alamos, NM 87545, USA}
\email{zmedin@lanl.gov}
\and
\author{Andrew Cumming}
\affil{Department of Physics, McGill University, 3600 rue University, Montreal, QC, H3A2T8, Canada}
\email{cumming@physics.mcgill.ca}

\begin{abstract}
We discuss the effect of chemical separation as matter freezes at the
base of the ocean of an accreting neutron star, and the subsequent
enrichment of the ocean in light elements and inward transport of heat
through convective mixing. We extend the steady-state results of
\cite{medin11} to transiently accreting neutron stars, by considering
the time-dependent cases of heating during accretion outbursts and
cooling during quiescence. Convective mixing is extremely efficient,
flattening the composition profile in about one convective turnover
time (weeks to months at the base of the ocean). During accretion
outbursts, inward heat transport has only a small effect on the
temperature profile in the outer layers until the ocean is strongly
enriched in light elements, a process that takes hundreds of years to
complete. During quiescence, however, inward heat transport rapidly
cools the outer layers of the ocean while keeping the inner layers
hot. We find that this leads to a sharp drop in surface emission at
around a week followed by a gradual recovery as cooling becomes
dominated by the crust. Such a dip should be observable in the light
curves of these neutron star transients, if enough data is taken at a
few days to a month after the end of accretion. If such a dip is
definitively observed, it will provide strong constraints on the
chemical composition of the ocean and outer crust.
\end{abstract}


%%%%
\section{Introduction}
\label{sec:intro}

The outermost $\simeq 30$~m of an accreting neutron star is expected
to form a fluid ocean that overlies the kilometer-thick solid crust of
the star \citep{bildsten95}. This ocean is of interest as the site of
long duration thermonuclear flashes such as superbursts
\citep{cumming01,strohmayer02,kuulkers04} and intermediate duration
bursts \citep{intzand05,cumming06}, non-radial oscillations
\citep{bildsten95,piro05} and because the matter in the ocean
eventually solidifies as it is compressed to higher densities by
ongoing accretion, and so determines the thermal, mechanical and
compositional properties of the neutron star crust
\citep{haensel90,brown98,schatz99}. The thermal properties of the
ocean determine the initial cooling of an accreting neutron star
following the onset of quiescence \citep[][hereafter BC09]{brown09} as
observed for 6 sources
\citep{wijnands01,wijnands02,wijnands03,wijnands04,cackett06,homan07,cackett08,fridriksson11,degenaar11a,degenaar11b,degenaar13b,cackett13}.

The composition of the ocean is expected to consist of mostly heavy
elements, formed by rapid proton capture (the rp-process) during
nuclear burning of the accreted hydrogen and helium at low densities
and subsequent electron captures at higher densities, although some
carbon may also be present \citep{schatz01,gupta07}. At the
ocean-crust boundary, as the matter transitions from liquid to solid
it also undergoes chemical separation. Numerical simulations of phase
transitions in neutron stars \citep*{horowitz07} have shown that as
the ocean mixture solidifies, the lighter elements (charge numbers $Z
\alt 20$) are preferentially left behind in the liquid whereas the
heavier elements are preferentially incorporated into the solid. In
\citet{medin11} (hereafter Paper~I), we showed that the retention of
light elements in the liquid acts as a source of buoyancy that drives
a continual mixing of the ocean, enriching it substantially in light
elements and leading to a relatively uniform composition with
depth. Heat is also transported inward to the ocean-crust boundary by
this convective mixing. In \citet{medin14} (hereafter Paper~II) we
showed that during quiescence the inward heat transport is
particularly strong, leading to rapid cooling of the outer ocean and a
significant drop in the light curve compared with standard cooling
models (e.g., BC09).

One motivation for studying the problem of ``compositionally driven''
convection in the neutron star ocean comes from superbursts, which are
thought to involve thermally unstable carbon burning in the deep ocean
of the neutron star \citep{cumming01,strohmayer02}. The energy release
in these very long duration thermonuclear flashes, inferred from
fitting their light curves \citep{cumming06}, corresponds to carbon
fractions of $\approx 20\%$. This has been challenging to produce in
models of the nuclear burning of the accreted hydrogen and helium. If
the hydrogen and helium burn unstably, the amount of carbon produced
is $\alt 1\%$ \citep{woosley04}, and whereas stable burning can
produce large carbon fractions \citep{schatz03,stevens14},
time-dependent models do not show stable burning at the $\approx 10\%$
Eddington accretion rates of superburst sources \citep[although
observationally, superburst sources show evidence that much of the
accreted material may not burn in Type I
bursts;][]{intzand03}. Perhaps even more problematic than making
enough carbon is that carbon ignition models for superbursts require
large ocean temperatures $\approx 6\times10^8~{\rm K}$ at the ignition
depth, which are difficult to achieve in standard models of crust
heating \citep[e.g.,][]{brown04,cumming06,keek08}.

Observations of quiescent transiently accreting neutron stars also
provide strong motivation for studying ocean convection. BC09 inferred
a large inward heat flux in the outer crust of the sources
MXB~1659--29 and KS~1731--260 by fitting their light curves in
quiescence. Other anomalous behavior from transiently accreting
neutron stars includes a rebrightening during a cooling episode in
XTE~J1701--462 \citep{fridriksson11}, and very rapid cooling a few
days after accretion shut off in XTE~J1709--267
\citep{degenaar13b}. Though both the rebrightening and the rapid
cooling can be explained by a spurt of accretion during quiescence, as
we show here these features may naturally arise from the cooling ocean
when chemical separation occurs.

In this paper we generalize and expand on the results of Papers~I and
II: We place the steady-state calculations of Paper~I in a larger
context by adding the relevant physics into a full
envelope-ocean-crust model (cf.\ \citealt{brown04}; BC09; Paper~II)
and by considering the evolution toward that steady state; and we
examine the quiescence calculations of Paper~II in greater detail and
provide cooling curve fits for several additional sources. We begin in
Section~\ref{sec:convection} by reviewing the picture of steady-state,
compositionally driven convection as presented in Paper~I, and discuss
how the picture changes when time dependence is considered. In
Section~\ref{sec:model} we describe our calculation of the
time-dependent temperature and composition structure of the ocean and
crust. In Sections~\ref{sec:accretion} and \ref{sec:cooling} we
present results from our calculation during accretion and during
cooling after accretion turns off, respectively; in
Section~\ref{sub:analytic} we additionally provide an analytic
approximation to our cooling model. In Section~\ref{sec:observation}
we compare the cooling light curves we generate to observations of
transiently accreting neutron stars during quiescence. Finally, in
Section~\ref{sec:discuss} we discuss the implications of our results.

%%%%
\section{Compositionally driven convection in the ocean}
\label{sec:convection}

\subsection{Phase diagrams and chemical separation}
\label{sub:phased}

As in Paper~I, to understand the effect of compositionally driven
convection on the ocean we first consider the phase diagram for the
ocean mixture. Though the ocean in an accreting neutron star is likely
made up of a wide variety of elements \citep{schatz01,gupta07}, for
computational tractability we only consider a two-component mixture of
oxygen and selenium in this paper. Our O-Se mixture is modeled after
the 17-component, rp-process ash mixture considered by
\cite{horowitz07} \citep[see also][]{gupta07}; in that latter mixture
selenium is the most abundant element and oxygen is the most abundant
low-$Z$ element. While the relative abundances and mass numbers of
each element change with depth due to electron captures, we ignore any
such effects and use ${}^{16}$O-${}^{79}$Se throughout the ocean. It
is unclear whether including two components is enough to accurately
model the effects of chemical and phase separation in the ocean, and
if so, what the charge values of those two components should
be. Calculations of multicomponent phase diagrams using both
extrapolation \citep[cf.][]{medin10} and molecular dynamics techniques
\citep[e.g.,][]{hughto12} are in progress to address these
issues. Note that the equations in the body of the paper are specific
to two-component mixtures, but that unless otherwise specified the
equations in the Appendix are applicable generally to multicomponent
mixtures.

The Coulomb coupling parameter is an important quantity for
determining the phase diagram of two-component mixtures. The Coulomb
coupling parameter for ion species $i$ is
\be
\Gamma_i = \frac{Z_i^{5/3} e^2}{k_BT} \left(\frac{4\pi\rho Y_e}{3m_p}\right)^{1/3} \,,
\ee
while that for the mixture is
\bal
\Gamma = {}& \frac{\langle Z^{5/3} \rangle e^2}{k_BT} \left(\frac{4\pi\rho Y_e}{3m_p}\right)^{1/3} \nonumber\\
 = {}& 204~\rho_9^{1/3}\left(\frac{T_8}{3}\right)^{-1}\left(\frac{\langle Z^{5/3} \rangle}{357}\right)\left(\frac{Y_e}{0.43}\right)^{1/3} \,.
\label{eq:gamma}
\eal
Here $Z$ and $A$ are the ion charge and mass number, $Y_e = \langle Z
\rangle/\langle A \rangle$ is the electron fraction, $\rho_9 =
\rho/(10^9~{\rm g~cm^{-3}})$ the density, and $T_8 = T/(10^8~{\rm K})$
the temperature; $\langle Q \rangle$ signifies the number average of
quantity $Q$ for the mixture, such that $\langle Q \rangle = x_1 Q_1 +
x_2 Q_2$, where $x_i$ is the number fraction of species $i$.

\begin{figure}
\begin{center}
\includegraphics[width=\columnwidth]{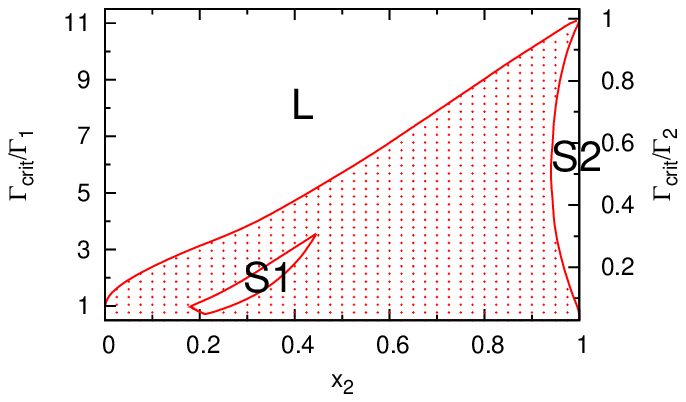}
\includegraphics[width=\columnwidth]{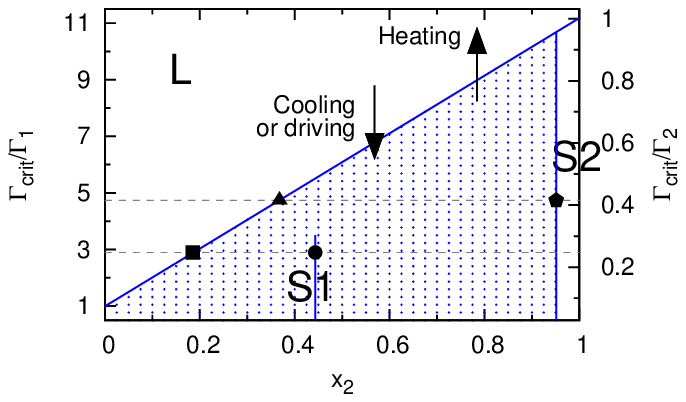}
\end{center}
\caption[The phase diagram]
{The phase diagram for crystallization of a two-component mixture with
charge ratio $Z_2/Z_1 = 34/8$ (top panel), and its approximation as
used in our simulation (bottom panel). The Coulomb coupling constants
$\Gamma_1$ and $\Gamma_2$ are given in terms of $\Gamma_{\rm crit}
\approx 175$, the value at which a single-species plasma
crystallizes. The stable liquid region of each phase diagram is
labeled ``L'', the stable solid regions are labeled ``S1'' and ``S2'',
and the unstable region is filled with plus symbols. A particle in the
ocean/crust moves down the phase diagram during cooling or accretion
driving, and up during rapid heating. In the bottom panel, the liquid
composition marked by a filled triangle is in equilibrium with the
solid composition marked by a filled pentagon; the liquid composition
marked by a filled square is in equilibrium with the solid composition
marked by a filled circle.}
\label{fig:phased}
\end{figure}

Figure~\ref{fig:phased} shows two phase diagrams for a two-component
mixture with charge ratio $Z_2/Z_1 = 34/8 = 4.25$, appropriate for,
e.g., an oxygen-selenium mixture (charge ratio $4.25$). The top panel
shows the detailed phase diagram calculated in \cite{medin10}; the
bottom panel shows the approximate phase diagram used in the
calculations in this paper. In each panel, the x-axis shows $x_2$, the
number fraction of the heavier ion species, and the left y-axis shows
$\Gamma_1^{-1}$, the inverse of the Coulomb coupling parameter for the
lighter species. For reference, the right y-axis shows
$\Gamma_2^{-1}$, the inverse of the coupling parameter for the heavier
species. In addition, ``L'' denotes the stable liquid region, ``S1''
and ``S2'' denote stable solid regions, and the shaded region
represents the unstable region of the phase diagram. A parcel with
composition and temperature (or equivalently, $x_2$ and $\Gamma_1$)
that lies inside the unstable region will undergo phase separation,
separating into two phases with compositions on either side of the
unstable region.\footnote{In the special case that $\Gamma_1$ is at
its eutectic value, the parcel separates into three phases; for the
example of Fig.~\ref{fig:phased}, when $\Gamma_{\rm
crit}/\Gamma_1=3.6$, the parcel separates into L, S1, and S2. See
Paper~I.} In this way, chemical separation occurs.

For a single species of ion, solidification occurs when $\Gamma >
\Gamma_{\rm crit} \approx 175$ \citep[e.g.,][]{potekhin00}. For a
multicomponent mixture a liquid becomes unstable to phase separation
at a $\Gamma$ value that varies with composition, known as the
liquidus curve (in Fig.~\ref{fig:phased}, the upper boundary of the
unstable region). For the $Z_2/Z_1 = 4.25$ charge mixture shown in
Fig.~\ref{fig:phased} the liquidus curve is almost linear in
$\Gamma_1^{-1}$ vs. $x_2$, with $\Gamma \simeq \Gamma_{\rm crit}$; we
have therefore chosen $\Gamma = \Gamma_{\rm crit}$ as the liquidus
curve for our approximate phase diagram. Using Eq.~(\ref{eq:gamma})
and the equation of state of relativistic, degenerate electrons
(applicable for $\rho \agt 10^7~{\rm g~cm^{-3}}$)
\be
y = \frac{P}{g} = \frac{(3\pi^2)^{1/3}}{4} \frac{\hbar c}{g} \left(\frac{\rho Y_e}{m_p}\right)^{4/3} \,,
\ee
where $y$ is the column depth, $P$ is the pressure, and $g$ is the
surface gravity, we have that the liquidus in our phase diagram
corresponds to the column depth
\be
y_L = \left[\frac{3}{16}\left(\frac{9}{4\pi^2}\right)^{1/3} \frac{\hbar c}{g} \left(\frac{\Gamma_{\rm crit}k_BT}{\langle Z^{5/3} \rangle e^2}\right)^4\right]_L \,,
\label{eq:yL}
\ee
where $T$ and $\langle Z^{5/3} \rangle$ are evaluated at the
liquidus. Although a multicomponent liquid becomes unstable to phase
separation at the liquidus, in general it does not completely solidify
until a much larger value of $\Gamma$. However, we found in Paper~I
that in the neutron star ocean any liquid-solid mixture formed during
phase separation will differentiate spatially due to sedimentation of
the solid particles at a rate much faster than any of the other mixing
processes (including accretion driving).\footnote{This requires the
solid particles to be denser than the liquid, which for the phase
diagram shown in Fig.~\ref{fig:phased} is true at all compositions
except $x_1$ very close to unity.} This means that all of the liquid
in the ocean-crust region will lie above all of the solid there, such
that the liquid effectively solidifies at the liquidus; the liquidus
depth is also the depth of the ocean-crust boundary $y_b$. In other
words, from Eq.~(\ref{eq:yL})
\bal
{}& y_b \equiv y_L \\
{}& = 5.57\times10^{12} \left(\frac{T_{b,8}}{3}\right)^4\left(\frac{\langle Z_b^{5/3} \rangle}{357}\right)^{-4}\left(\frac{g_{14}}{2.45}\right)^{-1}~{\rm g~cm^{-2}} \,,
\label{eq:yb}
\eal
where $T_b$ and $Z_b$ are the temperature and ion charge at the base
of the ocean.

\subsection{Regimes of chemical separation}
\label{sub:regimes}

The fate of ocean-crust particles as they cross into the unstable
region of the phase diagram and undergo phase/chemical separation is
determined by the composition of the parcels before crossing and the
direction they are moving on the diagram. The initial composition of
the parcels depends on the accretion history of the neutron star. The
direction each parcel moves on the phase diagram depends on whether
accretion is ongoing or not; and if accretion is ongoing, whether the
rate at which the ocean-crust boundary moves inward is greater than
the rate at which particles are driven inward, i.e., whether
$\dot{y}_b > \dot{m}$, where $\dot{y}_b$ is the rate of change in
$y_b$ and $\dot{m}$ is the local accretion rate per unit area. There
are three regimes to consider: steady-state accretion, cooling after
accretion shuts off, and rapid heating shortly after accretion turns
on.

1) When the neutron star is accreting and the ocean-crust system is
near or at its steady-state configuration, $\dot{y}_b <
\dot{m}$. In this regime accreted material is driven to higher
pressure, such that material at the base of the ocean moves across the
ocean-crust boundary and freezes (as shown by the arrow marked
``driving'' in Fig.~\ref{fig:phased}). According to the simplified
phase diagram in Fig.~\ref{fig:phased}, if the ocean base has a
composition $x_2 > 0.95$ or $x_2 = 0$, there will be no chemical
separation upon freezing. If $0.95 > x_2 > 0.25$, some material will
remain liquid and some will form a solid of composition S2. If $0.25 >
x_2 > 0$, some material will remain liquid and some will form a solid
of composition S1.

2) When accretion turns off and the neutron star is cooling,
$\dot{y}_b < 0$ and $\dot{m} = 0$. In this regime the temperature
drops in the ocean, such that the ocean-crust boundary moves outward
and the base of the ocean freezes (``cooling'' in
Fig.~\ref{fig:phased}). In this case the behavior of chemical
separation with $x_2$ will be the same as that described above.

3) When accretion turns on again and the system moves toward its
steady-state configuration, the heating is initially very
strong. In this regime the crust melts faster than new material can
be driven across the ocean-crust boundary, such that $\dot{y}_b >
\dot{m}$ (and material follows the ``heating'' arrow in
Fig.~\ref{fig:phased}). According to our simplified phase diagram, if
the crust has a composition $x_2 = 0$, $x_2 > 0.95$, or S2, there is
no chemical separation of the solid upon melting. If the crust is of
composition S1, there will be chemical and phase separation into a
light liquid and an S2 solid. However, assuming diffusion between
solid-solid phases is slow \citep[cf.][]{hughto11}, as heating
continues the liquid and the solid will travel together up the phase
diagram until the solid melts and recombines with the liquid; since
the distance over which this occurs is relatively short compared to
the size of the ocean, we assume for simplicity that in our
calculations a solid of composition S1 will just melt to form a liquid
of composition S1 ($x_2 = 0.44$). Therefore, in our calculations there
is no chemical or phase separation when $\dot{y}_b > \dot{m}$,
regardless of crust composition.

\subsection{Compositional buoyancy and convection}
\label{sub:buoy}

What happens to the liquid left behind after phase separation and
sedimentation of the solid depends on its composition. During rapid
heating (regime~3 of the previous subsection) the liquid remains in
place at the base of the ocean, since it is heavier than or at the
same composition as the liquid above it. During steady-state accretion
or cooling (regimes~1 and 2), however, after the solid particles form
and sediment out the liquid left behind is lighter than the liquid
immediately above it and so will have a tendency to buoyantly
rise. This is counteracted by the thermal profile, which is stably
stratified in the absence of a composition gradient such that a rising
fluid element will be colder than its surroundings and will tend to
sink back down.

A measure of the buoyancy is the convective (Schwarzschild)
discriminant ${\mathcal A}$, which is related to the
Brunt-V\"ais\"al\"a frequency $N^2=-g{\mathcal A}$ \citep{cox80}. In
Appendix~\ref{sec:stable} we derive the equations for convective
stability; for a two-component mixture we can write
[Eq.~(\ref{eq:Aconv2}); see also \citealt{kippenhahn94}]
\be
{\mathcal A}H_P\chi_\rho = \chi_T\left(\nabla -\nabla_{\rm ad}\right) + \chi_1\nabla_{X_1} \,.
\label{eq:A}
\ee
Here,
\be
\chi_1 = \chi_{X_1}-\chi_{X_2}\frac{X_1}{X_2}+\chi_{Y_e}\frac{(Y_1-Y_2)X_1}{Y_e} \,,
\label{eq:chi1}
\ee
$X_i = x_i A_i/\langle A \rangle$ is the mass fraction of species $i$,
$Y_i = Z_i/A_i$ is the electron fraction of species $i$, $H_P =
-dr/d\ln y = y/\rho$ is the scale height, and $\nabla = -H_P(d\ln
T/dr)$ and $\nabla_{X_i} = -H_P(d\ln X_i/dr)$ are the temperature and
composition gradients. The adiabatic temperature gradient is taken at
constant (specific) entropy $s$ and composition $\{X_i\}$:
$\nabla_{\rm ad} = -H_P(d\ln T /dr)_{s,X_1,X_2,Y_e}$. For a quantity
$Q$, $\chi_Q = (\partial \ln P/\partial \ln Q)$ with the other
independent thermodynamic variables held constant. Although neither
$X_2$ nor $Y_e$ are independent variables, being subject to the
constraints $X_2 = 1-X_1$ and $Y_e = Y_1X_1 + Y_2X_2$, here we treat
them as such in order to show explicitly the ion and electron
contributions to various expressions in this paper (e.g., the specific
heat given below). The ion and electron contributions are then
combined in Eqs.~(\ref{eq:chi1}) and (\ref{eq:b1}). Note that
$\chi_1$, $\chi_T$, and $\chi_\rho$ are all positive quantities. If
${\mathcal A} < 0$ or $N^2 > 0$ the ocean is stable to convection. For
example, if the composition is uniform so that $\nabla_{X_1} = 0$,
stability to convection requires the familiar condition $\nabla <
\nabla_{\rm ad}$. The maximum value of $\chi_1\nabla_{X_1}$ such that
the ocean is stable to convection is therefore
$\chi_T\left(\nabla_{\rm ad}-\nabla\right)$.

As steady-state accretion or cooling continues, light elements are
continually deposited at the base of the ocean and must be transported
upwards by convection. For efficient convection $\cal{A}$ adjusts to
be close to but slightly greater than zero. In Paper~I we found that
convection is extremely efficient throughout the ocean during
steady-state accretion; in Appendix~\ref{sec:mixing} of this paper we
demonstrate that convection in the ocean is extremely efficient even
during time-dependent heating or cooling, and even when effects due to
rapid rotation ($\sim 10^2~{\rm s^{-1}}$) and moderate magnetic fields
($\sim 10^{10}$~G) are considered. We therefore assume in the main
body of the paper that
\be
\chi_1\nabla_{X_1} = \chi_T\left(\nabla_{\rm ad}-\nabla\right)
\label{eq:nablaXe}
\ee
where convection is active (i.e., across the ocean convection zone).

\subsection{Convection equations}
\label{sub:equations}

Here we assume Newtonian physics, plane-parallel geometry, mixing
length theory, and efficient convection [Eq.~(\ref{eq:nablaXe})]. In
mixing length theory the value of the mixing length is highly
uncertain; but note below that under the efficient convection
assumption this parameter does not appear in our equations. The
continuity equation for the flow of species $i$ is given by
\be
\frac{\partial X_i}{\partial t} + \dot{m}\frac{\partial X_i}{\partial y} = \frac{\partial F_{r,X_i}}{\partial y} + \epsilon_{X_i} \,,
\label{eq:xflux}
\ee
where $\mathbf{F}_{X_i} = F_{r,X_i} \hat{\mathbf{r}}$ is the
composition flux for species $i$ and $\epsilon_{X_i}$ is the sum of
all composition sources. The entropy balance equation is given by
\citep[e.g.,][]{brown98}
\be
T\frac{\partial s}{\partial t} + T\dot{m}\frac{\partial s}{\partial y} = \frac{\partial F_r}{\partial y} + \epsilon \,,
\label{eq:EBE}
\ee
where
\be
\mathbf{F} = F_r \hat{\mathbf{r}} = \mathbf{F}_{\rm cd} + \mathbf{F}_{\rm conv}
\label{eq:Fr}
\ee
is the total flux,
\be
\mathbf{F}_{\rm cd} = F_{r,\rm cd} \hat{\mathbf{r}} = \frac{KT}{H_P}\nabla \hat{\mathbf{r}}
\label{eq:Fcd}
\ee
is the conductive heat flux, $\mathbf{F}_{\rm conv} = F_{r,\rm conv}
\hat{\mathbf{r}}$ is the convective heat flux, $K$ is the thermal
conductivity, and $\epsilon$ is the sum of all heat sources. The terms
on the left-hand side of Eq.~(\ref{eq:EBE}) can be written
[Eqs.~(\ref{eq:dsdt}) and (\ref{eq:dsdr})]
\be
T\frac{\partial s}{\partial t} = c_P \frac{\partial T}{\partial t} - \frac{b_1T}{X_1}\frac{\partial X_1}{\partial t}
\label{eq:Tdsdt}
\ee
and
\be
T\dot{m}\frac{\partial s}{\partial y} = \frac{c_PT\dot{m}}{y}\left(\nabla - \nabla_{\rm ad} - \frac{b_1}{c_P} \nabla_{X_1}\right) \,,
\label{eq:Tmdsdy}
\ee
where $c_P = T(\partial s/\partial T)_{P,X_i,Y_e}$ is the specific
heat capacity,
\be
b_1 = b_{P,1}-b_{P,2}\frac{X_1}{X_2}+b_{P,e}\frac{(Y_1-Y_2)X_1}{Y_e} \,,
\label{eq:b1}
\ee
$b_{P,i} = X_i(\partial s/\partial X_i)_{T,P,X_{j \ne i},Y_e}$, and
$b_{P,e} = Y_e(\partial s/\partial Y_e)_{T,P,X_1,X_2}$. With the
assumption of efficient convection, the convective heat flux in the
ocean becomes [Eq.~(\ref{eq:AeF})]
\be
F_{r,\rm conv} = -\frac{c_PT\chi_1}{X_1\chi_T} \left(1 + \frac{\chi_Tb_1}{\chi_1c_P}\right) F_{r,X_1} \,,
\label{eq:eF}
\ee
the entropy balance equation in the ocean becomes
[Eq.~(\ref{eq:AEBE2})]
\bal
c_P\frac{\partial T}{\partial t} + \frac{c_PT\chi_1}{X_1\chi_T}\frac{\partial X_1}{\partial t} \nonumber\\
 = \frac{\partial F_{r,\rm cd}}{\partial y} - F_{r,X_1} \frac{\partial}{\partial y} \left[\frac{c_PT\chi_1}{X_1\chi_T} \left(1 + \frac{\chi_Tb_1}{\chi_1c_P}\right)\right] + \epsilon \,,
\label{eq:EBE2}
\eal
and the entropy advection term in the ocean becomes
\be
T\dot{m}\frac{\partial s}{\partial y} = -\frac{c_PT\dot{m}\chi_1}{y\chi_T}\left(1 + \frac{\chi_Tb_1}{\chi_1c_P}\right)\nabla_{X_1} \,.
\label{eq:Tmdsdy2}
\ee

The steady-state versions of the above equations are similar to the
convection equations from Paper~I. Using Eq.~(\ref{eq:eF}) with the
steady-state composition flux $F_{r,X_i} = \dot{m}(X_i-X_{i,0})$
[Paper~I or Eq.~(\ref{eq:xfluxy})], we have that the steady-state
convective flux is given by
\be
F_{r,\rm conv} = -\frac{c_PT\dot{m}\chi_1}{\chi_T} \left(1 + \frac{\chi_Tb_1}{\chi_1c_P}\right)\left(1-\frac{X_{1,0}}{X_1}\right) \,.
\label{eq:Fconvss}
\ee
This equation differs from equation 43 of Paper~I (which is in error)
by the factor $1 + (\chi_T/\chi_1)(b_1/c_P)$, which is less than $1.2$
in any part of the ocean; the extra factor does not qualitatively
change the results of our earlier paper. In the deep ocean, because
$c_P$, $T$, $\chi_1$, $X_1$, and $b_1$ have only a weak dependence on
$y$ but $\chi_T \propto E_F \propto y^{-1/4}$, we also have that
\be
F_{r,\rm conv} \propto y^{1/4}
\label{eq:Fconvssprop}
\ee
(cf.\ the result during cooling, $F_{r,\rm conv} \propto y^{5/4}$; see
Paper~II) and
\be
\frac{\partial F_{r,\rm conv}}{\partial y} \simeq \frac{c_PT\dot{m}\chi_1}{4y\chi_T}\left(1-\frac{X_{1,0}}{X_1}\right) \,.
\label{eq:dFconvss}
\ee
Since $\nabla_{X_1} \ll 1$, we have from Eqs.~(\ref{eq:Tmdsdy2}) and
(\ref{eq:dFconvss}) that $T\dot{m}(\partial s/\partial y) \ll \partial
F_{r,{\rm conv}}/\partial y$. Therefore, using Eq.~(\ref{eq:EBE}) we
have that
\be
\frac{\partial F_r}{\partial y} \equiv \frac{\partial F_{r,\rm cd}}{\partial y} + \frac{\partial F_{r,\rm conv}}{\partial y} \simeq -\epsilon
\label{eq:EBEss}
\ee
in steady state, as we assumed in Paper~I.

%%%%
\section{A model of the envelope, crust, and ocean in the time-dependent case}
\label{sec:model}

In our model we place the top of the envelope at $y = 10^{-4}~{\rm
g~cm^{-2}}$ (i.e., at the surface) and the base of the crust at $y =
3\times10^{18}~{\rm g~cm^{-2}}$. The envelope structure is found as in
\citealt{brown02} \citep[see also][]{potekhin99}. We assume an
$\{X_{\rm H},X_{\rm He}\}=\{0.7,0.3\}$ composition throughout the
envelope. The crust structure and composition is found as in BC09,
except that we leave the core temperature $T_c$ as a free parameter
rather than solving for it self-consistently, and use Newtonian
physics with a surface gravity $g$ constant across the envelope,
ocean, and crust. General relativistic corrections are included only
as an overall redshift of the time, $t_\infty = t(1+z_{\rm surf})$,
and effective temperature, $T_{\rm eff,\infty} = T_{\rm eff}/(1+z_{\rm
surf})$, from the local value to that seen by an observer at infinity;
here the neutron star mass and radius are 1.62~$M_\sun$ and 11.2~km,
giving a redshift factor of $1+z_{\rm surf} = 1.32$. Note that while
$g$ varies by about ten percent across the crust of a neutron star, it
varies by less than a percent across the ocean, such that the
assumption of constant $g$ will modify the crust structure somewhat
but will have very little effect on the ocean structure (for a given
heat flux coming from the crust). As in BC09, we characterize the
thermal conductivity in the inner crust with a single number, the
impurity parameter $Q_{\rm imp} = \langle Z^2 \rangle - \langle Z
\rangle^2$. Our treatment of $\epsilon_{X_i}$ in the crust, as well as
our treatment of $\epsilon$ across all layers, is described in
Appendix~\ref{sec:sources}. In the ocean the components of $b_i$ are
found from the thermodynamic equations of Appendix~\ref{sec:thermo};
$K$, $c_P$, and the other thermodynamic derivatives are found as in
Paper~I (see, e.g., equation~39 of that paper).

\begin{figure}
\begin{center}
\includegraphics[width=\columnwidth]{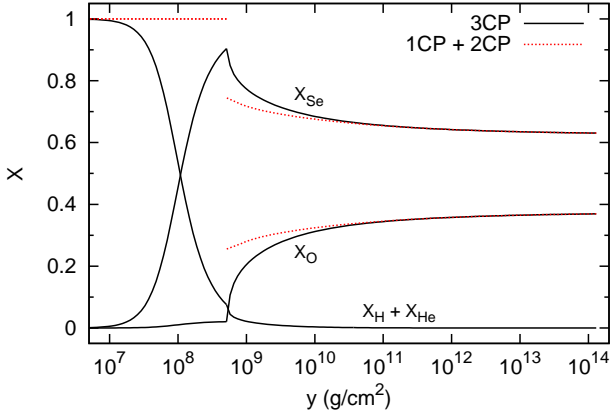}
\end{center}
\caption[The composition profile, with a H/He burning layer]
{Composition profiles in an ${}^{16}$O-${}^{79}$Se ocean and overlying
hydrogen-helium burning layer. The hydrogen and helium are treated as
a single component. The results from two models are shown: the
three-component plasma (``3CP'') model from Paper~I, where the burning
layer and the ocean are allowed to mix through convection and nuclear
reactions are included in a crude manner; and the one-component
burning layer and two-component ocean (``1CP + 2CP'') model from this
paper, where the two layers do not mix and nuclear reactions are
included only in the sense that $\left\{X_i\right\}$ changes at the
boundary ($y_0 = 5.1\times10^8~{\rm g/cm^2}$ for this example).}
\label{fig:Xburning}
\end{figure}

The ocean is bounded from above by the hydrogen and helium burning
layer, which ends at a column depth $y_0$. Rather than tracking the
physics of this layer, we leave $y_0$ as a free parameter; in
Sections~\ref{sec:accretion} and \ref{sec:cooling} we choose $y_0 =
10^8~{\rm g~cm^{-2}}$ (e.g., \citealt{bildsten97};
cf.\ Fig.~\ref{fig:Xburning}). The mass fraction of species $i$ at the
top of the ocean, $X_{i,0}$, is determined by the nuclear reactions
within the burning layer \citep[see][]{schatz01}. For the
${}^{16}$O-${}^{79}$Se mixture described in
Section~\ref{sec:convection} we nominally choose
$\{X_{1,0},X_{2,0}\}=\{0.02,0.98\}$, as in Paper~I (but see
below). This is approximately the mixture the ocean would have if all
of the light elements ($Z \le 20$) were oxygen and all of the heavy
elements ($Z > 20$) were selenium; further calculations, involving
mixtures of more than two components, are required to determine
whether this is a reasonable approximation.

The heavy-element ocean cannot penetrate into the
light-element burning layer, such that the burning layer is stable to
convection and the convective velocity drops to zero at the boundary
[Eq.~(\ref{eq:vconv})]. To include the stabilizing effect of this
layer in our model we set
\be
F_{r,X_i}(y_0) = 0
\label{eq:FX0}
\ee
and
\be
F_{r,\rm conv}(y_0) = 0 \,.
\label{eq:F0bc}
\ee
The boundary conditions Eqs.~(\ref{eq:FX0}) and (\ref{eq:F0bc}) can
occasionally be inconsistent with our assumption $X_{1,0} = 0.02$
above, in which case we allow $X_{1,0}$ to grow as
necessary. Figure~\ref{fig:Xburning} shows an example of a case where
$X_{1,0} > 0.02$ for our model (see also figure~3 from Paper~II). Note
that in the more accurate three-component model of Paper~I there is no
inconsistency between $F_{r,X_i}(y_0) = 0$ and $X_{1,0} = 0.02$,
because the required rapid rise in $X_{\rm O}$ with increasing column
depth is stabilized by the rapid drop in $X_{\rm H} + X_{\rm He}$. A
similar situation occurs when convection is thermally driven ($\nabla
\agt \nabla_{\rm ad}$ and $\sum_i \chi_i\nabla_{X_i}=0$) at the top of
the ocean.

The ocean is bounded from below by the crust, which begins at a column
depth $y_b$. The mass fraction of species $i$ at the top of the crust,
$X_{i,c}$, is determined by the mass fraction of species $i$ at the
base of the ocean, $X_{i,b}$, according to the relevant phase
diagram. For ${}^{16}$O-${}^{79}$Se we use Fig.~\ref{fig:phased} (see
also Section~\ref{sub:regimes}); converting from number fraction to
mass fraction gives
\be
X_{1,c} = \left\{
\begin{array}{ll}
X_{1,b} \,, & X_{1,b} \le 0.01~\mbox{or}~X_{1,b} = 1 \,; \\
0.01~\mbox{(``S2'')} \,, & 0.01 < X_{1,b} < 0.37 \,; \\
0.2~\mbox{(``S1'')} \,, & 0.37 < X_{1,b} < 1
\end{array}
\right.
\label{eq:X1c}
\ee
for $\dot{y}_b < \dot{m}$, and
\be
X_{1,c} = X_{1,b}
\ee
for $\dot{y}_b > \dot{m}$.

Convection can not occur for $y > y_b$, since the region is solid;
therefore, at the ocean-crust boundary we set
\be
F_{r,\rm conv}(y_b^+) = 0 \,,
\ee
where the superscript `$+$' signifies that the flux is evaluated on the
deep (i.e., crust) side of the boundary. The composition flux at the
ocean base is
\be
F_{r,X_i}(y_b^-) = (\dot{m}-\dot{y}_b) \Delta X_{i,bc} \,,
\label{eq:xfluxb}
\ee
where $\Delta X_{i,bc} = X_{i,b}-X_{i,c}$ and the superscript `$-$'
signifies that the flux is evaluated on the shallow (i.e., ocean) side
of the boundary [cf.\ the steady-state accretion expression
$\dot{m}(X_{i,b}-X_{i,0})$ of Paper~I]. If $\dot{y}_b \ge \dot{m}$,
there will be no chemical separation at the boundary
(Section~\ref{sub:regimes}) and therefore no compositionally driven
convection in the ocean, such that
\be
F_{r,X_i} = 0
\label{eq:FXnoc}
\ee
and
\be
F_{r,\rm conv} = 0
\label{eq:Fnoc}
\ee
throughout the ocean. Note that from Eqs.~(\ref{eq:X1c}) and
(\ref{eq:xfluxb}), $F_{r,X_1}(y_b^-) > 0$ for the O-Se system; with
Eq.~(\ref{eq:eF}) this means that $F_{r,\rm conv}(y_b^-) < 0$ or that
there is an inward heat flux at the base of the ocean due to
compositionally driven convection (cf.\ Paper~I).

We use a stationary grid for all column depths except $y_b$, which we
track continuously. The rate at which the ocean-crust boundary moves
is constrained by the heat flux continuity condition
\be
F_{r,\rm cd}(y_b^-) + F_{r,\rm conv}(y_b^-) = F_{r,\rm cd}(y_b^+) \,,
\label{eq:Fbbc}
\ee
which using Eqs.~(\ref{eq:eF}) and (\ref{eq:xfluxb}) becomes
\bal
\frac{c_PT_b(\dot{m}-\dot{y}_b)\chi_1}{\chi_T} \left(1 + \frac{\chi_Tb_1}{\chi_1c_P}\right)\left(1-\frac{X_{1,c}}{X_{1,b}}\right) \nonumber\\
 = F_{r,\rm cd}(y_b^-)-F_{r,\rm cd}(y_b^+) \,.
\label{eq:dotyb}
\eal
To estimate $F_{r,\rm cd}(y_b^-)$ and $F_{r,\rm cd}(y_b^+)$ we use the
temperature gradient between $y_b$ and the nearest grid point on the
low-$y$ side and on the high-$y$ side, respectively; for sufficiently
small grid spacing this is a reasonable approximation. During rapid
heating this approximation gives $F_{r,\rm cd}(y_b^-) < F_{r,\rm
cd}(y_b^+)$ or $\dot{y}_b > \dot{m}$, such that to maintain
self-consistency between Eqs.~(\ref{eq:Fnoc}) and (\ref{eq:Fbbc}) we
can not use Eq.~(\ref{eq:dotyb}) to find $\dot{y}_b$ but must find
$y_b$ from Eq.~(\ref{eq:yb}).

In this paper we assume for simplicity that $\epsilon_{X_i} = 0$ in
the ocean. In particular, this means that we ignore the effect of
electron captures on the ocean composition. Therefore, using
Eqs.~(\ref{eq:xflux}) and (\ref{eq:FX0}), the composition flux at any
point in the ocean satisfies
\be
F_{r,X_i}(y) = \int_{y_0}^y \frac{\partial X_i(y')}{\partial t} dy' + \dot{m}(X_i-X_{i,0}) \,.
\label{eq:xfluxy}
\ee
From Eqs.~(\ref{eq:xfluxb}) and (\ref{eq:xfluxy}) we have that the
total change in ocean composition with time is given by
\be
\int_{y_0}^{y_b} \frac{\partial X_i(y')}{\partial t} dy' = \dot{m}\Delta X_{i,0c} - \dot{y}_b\Delta X_{i,bc}
\label{eq:Xinout}
\ee
with $\Delta X_{i,0c} = X_{i,0}-X_{i,c}$; this expression is used as a
consistency check when we solve for $\partial X_1/\partial t$
below. The first term in Eq.~(\ref{eq:Xinout}) represents the balance
between the composition $\{X_{i,0}\}$ entering the ocean from the
burning layer and the composition $\{X_{i,c}\}$ leaving the ocean to
the crust, as driven by accretion; the second term represents the
exchange of particles in the ocean to convert a solid block of
composition $\{X_{i,c}\}$ into a liquid block of composition
$\{X_{i,b}\}$, as the boundary moves inward (or vice versa as the
boundary moves outward).

For the two-component ocean mixture considered in this paper, we solve
for the evolution of the ocean composition and temperature structure
as follows: In each time step we first guess a value for $\partial
X_{1,b}/\partial t$. Our guess comes from the fact that composition
changes slowly with depth near the base of the ocean, i.e.,
$\nabla_{X_1} = \chi_T\left(\nabla_{\rm ad}-\nabla\right)/\chi_1 \ll
1$; which with Eq.~(\ref{eq:Xinout}) gives the approximation
\be
\frac{\partial X_{1,b}}{\partial t} \simeq \frac{\dot{m}\Delta X_{1,0c} - \dot{y}_b\Delta X_{1,bc}}{y_b} \,,
\label{eq:Xbguess}
\ee
where $\dot{y}_b$ is obtained from Eq.~(\ref{eq:dotyb}). Once
$\partial X_{1,b}/\partial t$ is chosen, the update value
$\hat{X}_{1,b}$ is found from
\be
\hat{X}_{1,b} = X_{1,b} + \Delta t \frac{\partial X_{1,b}}{\partial t} \,,
\label{eq:dXdtexp}
\ee
where $\Delta t$ is the current time step; $\hat{X}_1$ at every other
depth in the ocean is found from $\hat{X}_{1,b}$ and
Eq.~(\ref{eq:nablaXe}), and then Eq.~(\ref{eq:dXdtexp}) is used to
find $\partial X_1/\partial t$ at each depth.  Finally,
Eq.~(\ref{eq:xfluxy}) is used with $\partial X_1/\partial t$ to find
$F_{r,X_1}$ for Eq.~(\ref{eq:EBE2}). The value of $\partial
X_{1,b}/\partial t$ is modified and the procedure repeated until
Eq.~(\ref{eq:Xinout}) holds true. The new value of $T_b$ is found from
$X_{1,b}$, $y_b$, and Eq.~(\ref{eq:yb}). We find that our initial
guess, Eq.~(\ref{eq:Xbguess}), often requires no extra iterations for
reasonable accuracy.

From Eq.~(\ref{eq:Xinout}), the steady-state $\partial X_i/\partial t
= 0$ is reached when $\dot{y}_b = 0$ and $\Delta X_{i,0c} = 0$; that
is, when the ocean-crust boundary stops moving and, if accretion is
ongoing, when the composition at the top of the ocean is the same as
that at the top of the crust (cf.\ Paper~I). The latter condition
happens in our ${}^{16}$O-${}^{79}$Se simulations when $X_{1,b} \simeq
0.37$ [see Eq.~(\ref{eq:X1c})] through a simple feedback mechanism: if
$ X_{1,b} < 0.37$ at the ocean base, $\Delta X_{1,0c} > 0$ (the crust
has composition S2), such that $\partial X_1/\partial t > 0$ and
$X_{1,b}$ rises above 0.37; if $X_{1,b} > 0.37$, $\Delta X_{1,0c} < 0$
(the crust has composition S1), such that $\partial X_1/\partial t <
0$ and $X_{1,b}$ drops below 0.37; the composition of the ocean base
hovers around $X_{1,b} = 0.37$. In reality the eutectic nature of the
phase diagram at $X_{1,b} = 0.37$ will most likely cause the ocean
base to solidify in vertically lamellar sheets of alternating S1, S2
composition \citep[e.g.,][]{woodruff73}.

%%%%
\section{Convection during accretion}
\label{sec:accretion}

Here we evolve an example O-Se ocean from quiescence to steady state
after accretion turns on, using the time-dependent equations of
Sections~\ref{sec:convection} and \ref{sec:model} and assuming that
compositionally driven convection is active. For this example $\dot{m}
= 10^5~{\rm g~cm^{-2}~s^{-1}}$ ($\sim 1.1 \dot{m}_{\rm Edd}$), $T_c =
10^8$~K, $Q_{\rm imp} = 0$, and $y_0 = 10^8~{\rm g~cm^{-2}}$. For our
initial conditions at the start of accretion we choose $T^{\rm init} =
T_c$ throughout the crust and ocean and $X_1^{\rm init} = 0.01$ (i.e.,
``S2'' in Fig.~\ref{fig:phased}) throughout the ocean. The latter
assumption is made because, even though $X_{1,b}$ can be large after a
cooling episode, during the initial heating there is rapid inward
movement of the ocean-crust boundary but no chemical separation, such
that the bulk of the ocean has the same composition as the accreted
crust (Section~\ref{sub:regimes}; but see Section~\ref{sec:discuss}).

\begin{figure}
\begin{center}
\includegraphics[width=\columnwidth]{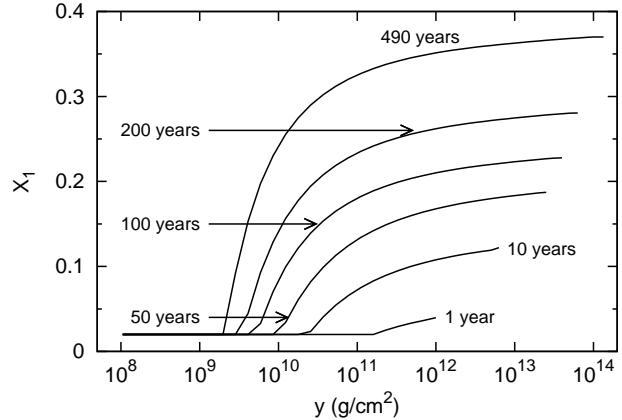}
\end{center}
\caption[The evolution of the composition profile during accretion]
{Composition profiles (in terms of the light element mass fraction
$X_1$) in the ocean of a neutron star with compositionally driven
convection, at various times after accretion turns on. Each curve is
labeled with a $t_\infty$ value; here $t_\infty$ is the time from the
start of the accretion outburst as seen by an observer at
infinity. Steady state is reached at $t_{\infty,\rm ss} = 490$~years,
or $t_{\rm ss} = t_{\infty,\rm ss}/(1+z_{\rm surf}) = 370$~years.}
\label{fig:Xevolve}
\end{figure}

Figure~\ref{fig:Xevolve} shows the composition profile at various
times during its evolution to steady state. The composition is shown
only for the ocean; the top of the ocean is located at $y = 10^8~{\rm
g~cm^{-2}}$, while the base of the ocean is located at the rightmost
extent of each curve and moves inward as $X_{1,b}$ increases. The top
of the convection zone can also be seen in the figure, as the depth
where $X_1$ reaches the burning layer level of 0.02 and flattens
out.

\begin{figure}
\begin{center}
\includegraphics[width=\columnwidth]{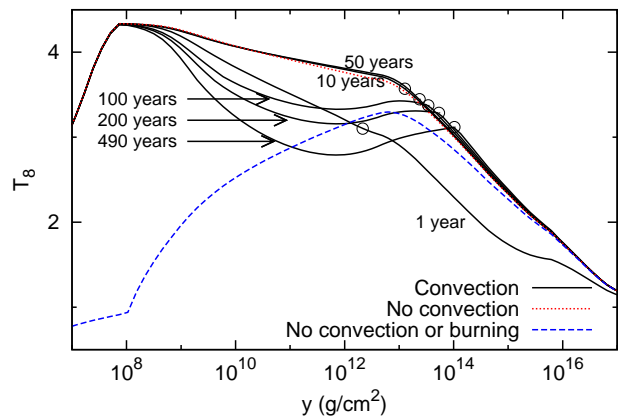}
\end{center}
\caption[The effect of convection on the steady-state temperature profile]
{The temperature profile in the outer layers of a neutron star with
compositionally driven convection, at various times after accretion
turns on. The model parameters are the same as in
Fig.~\ref{fig:Xevolve}. Each curve is labeled with a $t_\infty$ value,
and for each curve the location and temperature of the ocean-crust
boundary is marked with an open circle. The temperature profiles for
the cases without convection, and without convection or hydrogen and
helium burning, are also plotted for comparison. Note that steady
state for these latter cases is reached after only $t_\infty \simeq
15$~years.}
\label{fig:Tevolve}
\end{figure}

Figure~\ref{fig:Tevolve} shows the temperature profile at various
times during its evolution to steady state. As is discussed in
Section~\ref{sub:regimes}, the ocean moves through two regimes to
reach steady state: Initially there is no compositionally driven
convection, because of the strong accretion heating such that
$\dot{y}_b > \dot{m}$; the temperature profile reaches a quasi-steady
state that matches the steady-state profile in the case without
convection (the red dotted curve in Fig.~\ref{fig:Tevolve}) in only a
few years. Once that quasi-steady state is reached, $\dot{y}_b \ll
\dot{m}$ and the system slowly evolves over hundreds of years to the
final steady state (the black solid curve with $t_\infty =
490$~years).

While the ocean reaches the efficient convection state given by
Eq.~(\ref{eq:nablaXe}) relatively quickly, in approximately one
convective turnover time $t_{\rm conv} \sim 0.01 y_b/\dot{m}$ (months
to a few years; see Paper~I), it takes much longer to reach steady
state, as can be seen in Figs.~\ref{fig:Xevolve} and
\ref{fig:Tevolve}. The time from the start of the accretion outburst
to the start of steady state can be estimated from
Eq.~(\ref{eq:Xbguess}): for a steady-state composition at the base of
the ocean $X_{1,b}^{\rm ss} = 0.37$ (Section~\ref{sec:model}), and a
difference between the composition at the top of the ocean and the top
of the crust $\Delta X_{1,0c} = 0.01$ [Eq.~(\ref{eq:X1c})], we have
\be
t_{\rm ss} \simeq \frac{X_{1,b}^{\rm ss}}{\partial X_{1,b}/\partial t} \sim \frac{y_b}{\dot{m}}\frac{X_{1,b}^{\rm ss}}{\Delta X_{1,0c}} \sim 1000t_{\rm conv}
\label{eq:tss}
\ee
(i.e., hundreds to thousands of years). Using Eq.~(\ref{eq:tss}) with
$y_b \sim 10^{14}~{\rm g~cm^{-2}}$, we find a steady-state time of
$t_{\rm ss} \sim 10^3$~years (cf.\ Fig.~\ref{fig:Xevolve}). Note that
if we instead use $\dot{m} \sim 10^4~{\rm g~cm^{-2}~s^{-1}}$, as is
typical for low-mass X-ray binaries, $t_{\rm ss} \sim 10^4$~years. If
$y_b \sim 10^{13}~{\rm g~cm^{-2}}$, as in the model of
\cite{horowitz07} for $T_b \simeq 3\times10^8$~K, the time to reach
steady state is still large: $t_{\rm ss} \sim 10^2$~years. If the mass
fraction of light elements entering the ocean is ten times larger
($X_{1,0} \sim 0.2$), however, as is the case for stable hydrogen and
helium burning \citep[e.g.,][]{stevens14}, $t_{\rm ss} \sim
y_b/\dot{m}$ is of order the accretion time.

\begin{figure}
\begin{center}
\includegraphics[width=\columnwidth]{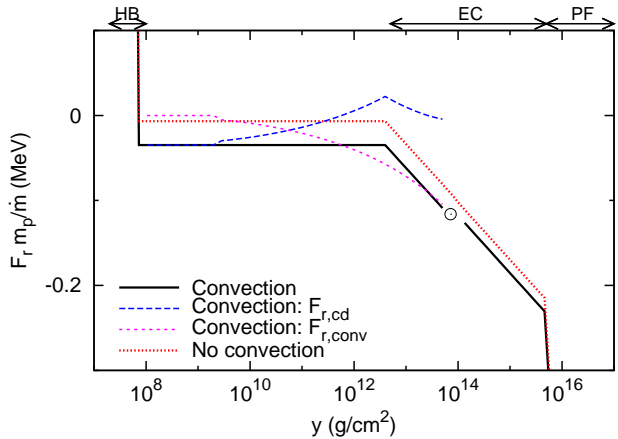}
\end{center}
\caption[The effect of convection on the steady-state flux]
{The total flux profile (in ${\rm MeV~nucleon^{-1}}$) in the outer
layers of a neutron star with compositionally driven convection, in
steady state and for the same parameters as in
Fig.~\ref{fig:Xevolve}. The location and flux of the ocean-crust
boundary is marked with an open circle; note also the gap at that
point, where we chose not to plot the flux due to the discontinuity in
the temperature derivative (Fig.~\ref{fig:Tevolve}). The labels that
appear above the graph denote the column depths where the various
accretion heat sources are active: HB is hydrogen and helium burning,
EC is electron captures, and PF is pycnonuclear fusion (see
Appendix~\ref{sec:sources}). The conduction ($F_{r,\rm cd}$) and
convection ($F_{r,\rm conv}$) flux profiles, as well as the total flux
profile for the case without convection, are also plotted for
comparison.}
\label{fig:Fprofile}
\end{figure}

Figure~\ref{fig:Fprofile} shows the steady-state profiles for the
total flux $F_r = F_{r,\rm cd} + F_{r,\rm conv}$ and the conduction
and convection contributions. The convection contribution has a
$y^{1/4}$ dependence, as in Eq.~(\ref{eq:Fconvssprop}). At the top of
the figure we have marked the locations of the three accretion heat
sources considered in this paper: hydrogen and helium burning,
electron captures, and pycnonuclear fusion (see
Appendix~\ref{sec:sources}). Since in steady state $\partial
F_r/\partial y \simeq -\epsilon$ [Eq.~(\ref{eq:EBEss})], $F_r$ is
constant outside of the heat source regions and drops by $\int_{y_{\rm
low}}^{y_{\rm high}} \epsilon dy$ across each region. For example,
electron captures are active from a column depth of $y_{\rm low} =
5\times10^{12}~{\rm g~cm^{-2}}$ to $y_{\rm high} = 5\times10^{15}~{\rm
g~cm^{-2}}$ and release a total energy of $Q_{\rm EC} = (m_p/\dot{m})
\int_{y_{\rm low}}^{y_{\rm high}} \epsilon dy = 0.2~{\rm
MeV~nucleon^{-1}}$ \citep[e.g.,][]{haensel08}, such that the total
drop in $F_r m_p/\dot{m}$ over that range is $0.2~{\rm MeV}$.

Note that the results shown in Figs.~\ref{fig:Tevolve} and
\ref{fig:Fprofile} are qualitatively different from those in figure~6
of Paper~I, despite the similarity in the parameters used. This is due
to a simplification made in our earlier paper: $F_{\rm crust}$, the
outward radial heat flux coming from the crust, is not the same for a
neutron star with compositionally driven convection as without. In
reality, $F_{\rm crust}$ must be solved self-consistently with $T_b$;
because $T_b$ is larger with convection, less heat flows into the
ocean from the crust and therefore $F_{\rm crust}$ is smaller (or more
negative, as is the case in Fig.~\ref{fig:Fprofile}), which limits the
growth of $T_b$ \citep[cf.\ figure~1 and discussion
of][]{brown04}. For the example of Fig.~\ref{fig:Tevolve} the
steady-state temperature at $y = 10^{14}~{\rm g~cm^{-2}}$ (near the
ocean crust-boundary) is only 4\% larger with convection than without;
whereas for the model of Paper~I it is 20\% larger. The inclusion of
hydrogen and helium burning has a comparable impact on our model,
increasing the temperature at $y = 10^{14}~{\rm g~cm^{-2}}$ by 5\%
(compare the blue dashed curve and the red dotted curve in
Fig.~\ref{fig:Tevolve}).

%%%%
\section{Convection after accretion turns off}
\label{sec:cooling}

We now consider the evolution of the ocean as it cools in
quiescence. We find that the evolution proceeds in four stages; these
stages are discussed in detail in Paper~II, but we outline them below
for reference (cf.\ Figs.~\ref{fig:Tcool} and \ref{fig:lightcurve}):

In stage~1, the base of the ocean has not yet started to cool and so
the evolution is the same with or without convection. In stage~2, the
cooling wave has reached the bottom of the ocean, and new crust begins
to form, driving convection. Inward heat transport by convection
rapidly cools the envelope and ocean but maintains the ocean-crust
boundary at a nearly constant temperature and depth. The temperature
gradient steepens with time. In stage~3, the temperature gradient at
the base of the ocean $\nabla_b$ reaches $\nabla_L\simeq 0.25$, the
liquidus temperature gradient. The region around the ocean-crust
boundary alternates between a state of strong convective heating and
crust melting, and a state of suppressed convection due to the release
of heavy elements into the ocean. The sporadic convection can no
longer prevent the ocean base from cooling, and cooling returns to a
level similar to during stage~1. In stage~4, the crust is thermally
relaxed, the ocean cools too slowly for convection to support the
steep gradient $\nabla_b = \nabla_L$, and the temperature profile in
the ocean flattens.

\subsection{Analytic approximation}
\label{sub:analytic}

Here we present an analytic approximation to the model of
Sections~\ref{sec:convection} and \ref{sec:model} applicable during
cooling. In our approximation, we assume that the ocean thermal
conductivity and the pressure scale height have the scaling
relationships
\be
K \propto y^{1/4} T \qquad{\rm and}\qquad H_P \propto y^{1/4} \,,
\label{eq:Kscale}
\ee
respectively; these relationships are valid when the electrons are
relativistic, i.e., for column depths around or greater than $y_t =
10^{10}~{\rm g~cm^{-2}}$. For each stage of cooling (see above or
Paper~II), we use Eq.~(\ref{eq:Kscale}) and a flux equation
[Eq.~(\ref{eq:Fcd1}) or (\ref{eq:Fcd2})] to solve for the temperature
$T_t$ at depth $y_t$; and then solve for the effective temperature
$T_{\rm eff}$ using the approximate relation (cf.\ BC09)
\be
\frac{d \ln T_{\rm eff}}{d \ln T_t} \simeq 0.5 \,,
\ee
or equivalently,
\be
T_{\rm eff} \simeq T_{\rm eff}^{(s)} \left(\frac{T_t}{T_t^{(s)}}\right)^{1/2} \,,
\label{eq:Teff}
\ee
where the superscript `(s)' signifies that the quantity is evaluated
at the beginning of stage~$s$ of cooling. To solve for the evolution
of $T_t$ we assume that the temperature profile through the ocean and
crust has an initially constant gradient $\nabla^{(1)}$
(cf.\ Fig.~\ref{fig:Tevolve}; see also below). During cooling, there
is a transition between the thermally relaxed outer layers with
constant outward heat flux $\sigma_B T_{\rm eff}^4$, where $\sigma_B$
is the Stefan-Boltzmann constant, and the inner layers still in steady
state with outward heat flux $KT\nabla^{(1)}/H_P$ [Eq.~(\ref{eq:Fcd})
with $\nabla = \nabla^{(1)}$]. While the cooling wave is still in the
ocean, this transition is defined by the thermal time
\be
\tau \simeq \frac{\rho c_P H_P^2}{2K} \propto yT^{-1}
\label{eq:tthermal}
\ee
\citep[cf.][]{henyey69}; the factor of two in Eq.~(\ref{eq:tthermal})
comes from integrating equation~7 of BC09 assuming
Eq.~(\ref{eq:Kscale}) for $K$ and $H_P$ and that $c_P$ and $T$ are
constant. We define the transition depth $y_\tau$ as the depth where
the thermal time is equal to the cooling time $t$.

During stage~1, the transition depth is above the base of the ocean,
i.e., $y_\tau < y_b^{(1)}$, and there is no compositionally driven
convection. The transition from the steady-state heat flux for $y >
y_\tau$ to the surface heat flux $\sigma_B T_{\rm eff}^4$ for $y \ll
y_\tau$ is not sharp (see, e.g., the ``20~days'' curve of
Fig.~\ref{fig:Tcool}). For lack of a better model, and to maintain
continuity between stages~1 and 2, we use a modified version of
Eq.~(\ref{eq:Fcd2}) for the heat flux: the (conductive) heat flux at
any point $y \le y_\tau$ is given by
\be
\frac{KT}{H_P}\nabla = \left[\left.\frac{KT}{H_P}\right|_{y=y_\tau}\nabla^{(1)} - \sigma T_{\rm eff}^4\right] \left(\frac{y}{y_\tau}\right)^{5/4} + \sigma T_{\rm eff}^4 \,.
\label{eq:Fcd1}
\ee
Here $Q|_{y=y_\tau}$ signifies that the quantity $Q$ is evaluated at
depth $y_\tau$. Equation~(\ref{eq:Fcd1}) has the desired properties of
being continuous and giving the correct heat flux values in the
limiting cases $y = y_\tau$ and $y \ll y_\tau$. With
Eq.~(\ref{eq:Kscale}) we can solve Eq.~(\ref{eq:Fcd1}) for the
temperature profile through the ocean,
\bal
T = {}& T_\tau \left[1 - \tfrac{8}{5}\left(\nabla^{(1)}-\nabla_{\rm eff,\tau}\right)\left\{1-\left(\frac{y}{y_\tau}\right)^{5/4}\right\} \right. \nonumber\\
 {}& \qquad\qquad\qquad \left. + 2\nabla_{\rm eff,\tau}\ln \left(\frac{y}{y_\tau}\right)\right]^{1/2} \,;
\eal
and the temperature at depth $y_t$ near the top of the ocean,
\be
T_t \simeq T_\tau \left[1-\tfrac{8}{5}(\nabla^{(1)}-\nabla_{\rm eff,\tau})+2\nabla_{\rm eff,\tau}\ln (y_t/y_\tau)\right]^{1/2} \,.
\label{eq:Tt1}
\ee
Here we assume $y_t \ll y_\tau$ and have defined
\be
\sigma T_{\rm eff}^4 \equiv \left.\frac{KT}{H_P}\right|_{y=y_\tau}\nabla_{\rm eff,\tau}
\label{eq:nablaeff1}
\ee
for convenience. From Eqs.~(\ref{eq:Teff}) and (\ref{eq:Tt1}) we
obtain the scaling relation
\bal
T_{\rm eff} = {}& T_{\rm eff}^{(2)} \left(\frac{T_\tau}{T_\tau^{(2)}}\right)^{1/2} \nonumber\\
 {}& \times \left[\frac{1-\tfrac{8}{5}\nabla^{(1)}}{1-\tfrac{8}{5}\nabla^{(1)}+2\nabla_{\rm eff,\tau}^{(2)}\ln \left(y_\tau/y_\tau^{(2)}\right)}\right]^{1/4} \,.
\label{eq:Teff1}
\eal
In deriving Eq.~(\ref{eq:Teff1}) we grouped $T_{\rm eff}$ terms and
used the fact that $\nabla_{\rm eff,\tau} \propto T_{\rm
eff}^4/T_\tau^2$ [Eq.~(\ref{eq:nablaeff1})].

To determine $T_\tau$ and $y_\tau$ as a function of time during
stage~1, we use Eq.~(\ref{eq:tthermal}):
\be
T_\tau = T_\tau^{(2)} \left(\frac{y_\tau}{y_\tau^{(2)}}\right)^{\nabla^{(1)}} = T_\tau^{(2)} \left(\frac{t}{t_2}\right)^{\nabla^{(1)}/\left(1-\nabla^{(1)}\right)}
\label{eq:Ttau1}
\ee
and therefore
\be
y_\tau = y_\tau^{(2)} \left(\frac{t}{t_2}\right)^{1/\left(1-\nabla^{(1)}\right)} \,,
\label{eq:ytau1}
\ee
where $t_2$ is the time at the beginning of stage~2 (see below). Along
with $T_\tau^{(2)} = T_b^{(2)}$, $y_\tau^{(2)} = y_b^{(2)}$, and
$\nabla_{\rm eff,\tau}^{(2)} = \nabla_{\rm eff}^{(2)}$ [where
$\nabla_{\rm eff}$ is $\nabla_{\rm eff,\tau}$ taken at the base of the
ocean; Eq.~(\ref{eq:nablaeff2})], Eqs.~(\ref{eq:Ttau1}) and
(\ref{eq:ytau1}) can be inserted into Eq.~(\ref{eq:Teff1}) to solve
for $T_{\rm eff}$ during stage~1 (cf.\ equation~8 of BC09).

During stages~2 and 3, the ocean is thermally relaxed, such that the
flux through the ocean satisfies $F_{r,\rm conv} + F_{r,\rm cd} =
\sigma_B T_{\rm eff}^4$. Since $F_{r,\rm conv} \propto y^{5/4}$
[Paper~I; see also Eq.~(\ref{eq:AeF}) with Eqs.~(\ref{eq:xfluxb}) and
(\ref{eq:Xbguess})] and $F_{r,\rm cd} = KT\nabla/H_P$, we have that
the conductive flux in the ocean is given by
\be
\frac{KT}{H_P}\nabla = \left[\left.\frac{KT}{H_P}\right|_{y=y_b}\nabla_b - \sigma T_{\rm eff}^4\right] \left(\frac{y}{y_b}\right)^{5/4} + \sigma T_{\rm eff}^4 \,.
\label{eq:Fcd2}
\ee
Similar to our method for stage~1 above, we use Eq.~(\ref{eq:Fcd2})
with Eq.~(\ref{eq:Kscale}) to solve for the temperature profile
through the ocean,
\bal
T = {}& T_b \left[1 - \tfrac{8}{5}\left(\nabla_b-\nabla_{\rm eff}\right)\left\{1-\left(\frac{y}{y_b}\right)^{5/4}\right\} \right. \nonumber\\
 {}& \qquad\qquad\qquad \left. + 2\nabla_{\rm eff}\ln \left(\frac{y}{y_b}\right)\right]^{1/2} \,,
\label{eq:T2}
\eal
and the temperature at depth $y_t$ near the top of the ocean,
\be
T_t \simeq T_b \left[1-\tfrac{8}{5}(\nabla_b-\nabla_{\rm eff})+2\nabla_{\rm eff}\ln (y_t/y_b)\right]^{1/2} \,,
\label{eq:Tt2}
\ee
where
\be
\sigma T_{\rm eff}^4 \equiv \left.\frac{KT}{H_P}\right|_{y=y_b}\nabla_{\rm eff}
\label{eq:nablaeff2}
\ee
and we assume that $y_t \ll y_b$. From Eqs.~(\ref{eq:Teff}) and
(\ref{eq:Tt2}) we obtain the scaling relation
\bal
T_{\rm eff} = {}& T_{\rm eff}^{(s)} \left(\frac{T_b}{T_b^{(s)}}\right)^{1/2} \nonumber\\
 {}& \times \left[\frac{1-\tfrac{8}{5}\nabla_b}{1-\tfrac{8}{5}\nabla_b^{(s)}+2\nabla_{\rm eff}^{(s)}\ln \left(y_b/y_b^{(s)}\right)}\right]^{1/4} \,.
\label{eq:Teff2}
\eal
In deriving Eq.~(\ref{eq:Teff2}) we grouped $T_{\rm eff}$ terms and
used the fact that $\nabla_{\rm eff} \propto T_{\rm eff}^4/T_b^2$
[Eq.~(\ref{eq:nablaeff2})].

Stage~2 begins at a time $t_2 = \tau_b^{(2)}$, where $\tau_b$ is the
thermal time evaluated at the base of the ocean. At the beginning of
this stage the ocean base is still in steady state, $\nabla_b^{(2)} =
\nabla^{(1)}$. We assume that the transition depth is stationary,
i.e., that $y_b = y_b^{(2)}$ and $T_b = T_b^{(2)}$ are constant. To
determine $\nabla_b$ as a function of time we look at the ocean
energetics: The total energy stored in the ocean is
\be
E = A \int_{y_t}^{y_b} c_P T dy \,,
\ee
where $A$ is the surface area; using Eq.~(\ref{eq:T2}) and assuming
that $c_P$ is constant in the ocean, that $y_t \ll y_b$, and that
$\nabla_b$ and $\nabla_{\rm eff}$ are small ($\nabla^{(1)} \le
\nabla_b \le \nabla_L$ in this stage and $\nabla_{\rm eff} \ll
\nabla_{\rm ad}$ typically), we have
\be
E \simeq A c_P T_b y_b \left(1 - \tfrac{4}{9}\nabla_b - \tfrac{5}{9}\nabla_{\rm eff}\right) \,.
\label{eq:Eapprox}
\ee
As $\nabla_b$ increases and the ocean cools, this energy is slowly
depleted; using Eq.~(\ref{eq:Eapprox}) and the fact that $y_b$ and
$T_b$ are constant during stage~2, we have that the ocean energy
changes at a rate
\be
\frac{dE}{dt} \simeq -A c_P T_b y_b \left(\frac{4}{9}\frac{d\nabla_b}{dt}+\frac{5}{9}\frac{d\nabla_{\rm eff}}{dt}\right) \,,
\label{eq:dEdt}
\ee
The depleted energy is released at the ocean base and must mask the
cooling due to the difference between the flux entering the ocean from
the crust and the flux leaving the ocean through the top; i.e.,
\be
-\frac{dE}{dt} = A\left.\frac{KT}{H_P}\right|_{y=y_b}\left(\nabla_{\rm eff} - \nabla^{(1)}\right) \,.
\label{eq:mdEdt}
\ee
From Eq.~(\ref{eq:Teff2}) we have that
\be
\frac{d\nabla_{\rm eff}}{dt} = -\frac{8\nabla_{\rm eff}^{(2)}}{5\left(1-\tfrac{8}{5}\nabla^{(1)}\right)}\frac{d\nabla_b}{dt} \,;
\label{eq:dnablaeff}
\ee
combining Eqs.~(\ref{eq:dEdt})--(\ref{eq:dnablaeff}) with $\nabla_{\rm
eff} \ll 1$ gives
\be
\nabla_b \simeq \tfrac{9}{8}\left(\nabla_{\rm eff}^{(2)}-\nabla^{(1)}\right)\left(\frac{t}{t_2}-1\right) + \nabla^{(1)} \,.
\label{eq:nablab}
\ee
Along with $T_b = T_b^{(2)}$ and $y_b = y_b^{(2)}$,
Eq.~(\ref{eq:nablab}) can be inserted into Eq.~(\ref{eq:Teff2}) to
solve for $T_{\rm eff}$ during stage~2.

Stage~3 begins when $\nabla_b = \nabla_L$, or at a time $t_3 = t_2
\left[8\left(\nabla_L-\nabla^{(1)}\right)/9\left(\nabla_{\rm
eff}^{(2)}-\nabla^{(1)}\right) + 1\right]$. We assume that $\nabla_b =
\nabla_L$ is constant. To determine $T_b$ and $y_b$ as a function of
time during stage~3, we use Eq.~(\ref{eq:tthermal}) with $c_P$ and $K$
at their solid values such that $\tau \propto y^{3/4}$ (BC09): Because
conduction is very efficient at transporting heat in the crust, we
assume that the temperature gradient in the crust from the ocean-crust
boundary to the transition depth is flat (cf.\ Paper~I); i.e., the
ocean-crust boundary cools at the same rate as the transition depth,
$\partial \ln T_b/\partial \ln t = \partial \ln T_\tau/\partial \ln t
= (\partial \ln T_\tau/\partial \ln y_\tau)(\partial \ln
y_\tau/\partial \ln \tau) = 4\nabla^{(1)}/3$, or equivalently,
\be
T_b = T_b^{(3)} \left(\frac{t}{t_3}\right)^{4\nabla^{(1)}/3} \,.
\label{eq:Tbanly}
\ee
If enrichment is low, $\partial \ln y_b/\partial t \simeq 4(\partial
\ln T_b/\partial t$) [Eq.~(\ref{eq:AdotT})], and we have
\be
y_b \simeq y_b^{(3)} \left(\frac{t}{t_3}\right)^{16\nabla^{(1)}/3} \,;
\label{eq:ybanly}
\ee
but if enrichment is high (as is the case in Fig.~\ref{fig:Tcool}; see
figure~3 of Paper~II), $\partial \ln y_b/\partial t \ll \partial \ln
T_b/\partial t$, and we have
\be
y_b \simeq y_b^{(3)} \,.
\label{eq:ybanly2}
\ee
Along with $\nabla_b = \nabla_L$, $T_b^{(3)} = T_b^{(2)}$, and
$y_b^{(3)} = y_b^{(2)}$, Eqs.~(\ref{eq:Tbanly})--(\ref{eq:ybanly2})
can be inserted into Eq.~(\ref{eq:Teff2}) to solve for $T_{\rm eff}$
during stage~3.

\subsection{Results}
\label{sub:results}

Here we evolve the O-Se ocean from Section~\ref{sec:accretion} as it
cools after accretion turns off. For stages~1, 2, and 4 of cooling we
use the equations from Sections~\ref{sec:convection} and
\ref{sec:model}, with $\dot{m} = 0$ and $\epsilon = 0$ as is
appropriate during cooling. We can also use these equations for
stage~3, but the resulting light curves are noisy unless the
simulation time step and spatial resolution are very small, due to the
quasi-periodic activation/deactivation of convection that occurs
during this stage (see above). Instead, we use the following method,
which has the advantage of requiring a much coarser time and spatial
grid for (empirically) comparably smooth and accurate light curves: We
assume that once compositionally driven convection is strong enough
for $\nabla_b = \nabla_L$, it will remain at that critical level as
cooling continues; i.e., we assume that when $\nabla_b \ge \nabla_L$,
\be
F_{r,\rm conv}(y_b^-) = F_{r,\rm cd}(y_b^+)-\frac{KT_b}{H_P}\nabla_L
\ee
[cf.\ Eq.~(\ref{eq:Fbbc})]. Equation~(\ref{eq:dotyb}) can no longer be
used to find $\dot{y}_b$, instead we use the ocean-crust boundary
equations of Appendix~\ref{sec:tracking}. From Eq.~(\ref{eq:Adotyb3})
we have [cf.\ Eq.~(\ref{eq:yb})]
\be
\dot{y}_b = \frac{4y_b}{T_b} \frac{\partial T_b}{\partial t} + y'_{b,1} \frac{\partial X_{1,b}}{\partial t} \,,
\label{eq:dotyb3}
\ee
where $y'_{b,1} = \partial y_b/\partial X_{1,b}-\partial y_b/\partial
X_{2,b}$. We solve for $\partial T_b/\partial t$ using the entropy
balance equation [Eq.~(\ref{eq:EBEb}); cf.\ Eq.~(\ref{eq:EBE})]
\bal
T_b\frac{\partial s_b}{\partial t} - T_b\dot{y}_b\left.\frac{\partial s}{\partial y}\right|_{y=y_b^-} = {}& \left.\frac{\partial F_{r}}{\partial y}\right|_{y=y_b} \\
 \simeq {}& \left[\frac{KT(\nabla_L-\nabla)}{H_P\Delta y}\right]_{y=y_b^-} \,,
\label{eq:Tdsdtb}
\eal
where $\Delta y$ is the grid spacing. Note that Eq.~(\ref{eq:Tdsdtb})
drives the temperature gradient $\nabla_b$ to $\nabla_L$. We solve for
$\partial X_{1,b}/\partial t$ using the iteration method described
earlier, except that our initial guess is [Eq.~(\ref{eq:AdotX})]
\be
\frac{\partial X_{1,b}}{\partial t} \simeq -\frac{4\Delta X_{1,bc}/T_b}{1+y'_{b,1}\Delta X_{1,bc}/y_b} \frac{\partial T_b}{\partial t} \,.
\ee
We use the above method whenever $\nabla_b > \nabla_L$ (i.e., during
stage~3), for all of the calculations shown here and in
Section~\ref{sec:observation}. Note that even with this method, the
light curves are slightly noisy in stage~3 (see, e.g.,
Fig.~\ref{fig:lightcurve}).

In this section we choose inital conditions at the start of cooling
$T^{\rm init}(y=10^{12}~{\rm g~cm^{-2}}) = 4\times10^8$~K near the
base of the ocean, $T_c^{\rm init} = 10^8$~K at the base of the crust,
and a constant temperature gradient in between; and $X_{1,b}^{\rm
init} = 0.37$ at the base of the ocean with a composition profile
given by Eq.~(\ref{eq:nablaXe}) throughout the ocean. These are
approximately the steady-state conditions from
Section~\ref{sec:accretion} (see also BC09; Paper~II). Note that our
assumption of an initially constant temperature gradient in the ocean
and crust means that the convective flux is zero at the start of
cooling, which is not entirely consistent with the steady-state
results of Section~\ref{sec:accretion}. Our intent here is to show the
effects of compositionally driven convection on cooling only. In
Section~\ref{sec:observation} we run our simulations over an entire
accretion cycle from outburst to quiescence, such that the convective
fluxes during cooling are calculated in a self-consistent way.

\begin{figure}
\begin{center}
\includegraphics[width=\columnwidth]{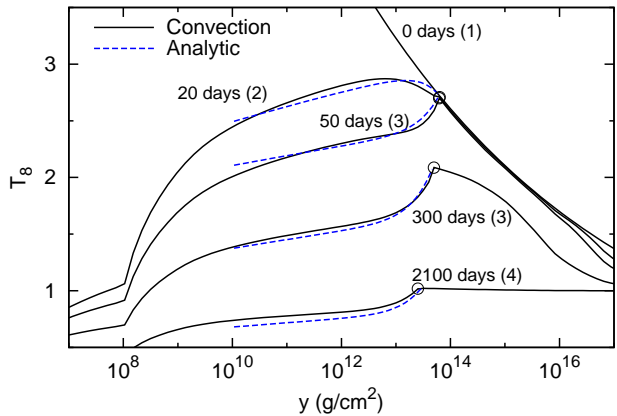}
\end{center}
\caption[The evolution of the temperature profile during cooling]
{Temperature profiles in the outer layers of a neutron star with
compositionally driven convection, and the analytic approximation to
those profiles [Eq.~(\ref{eq:T2})], at various times after accretion
turns off. Each curve is labeled with a $t_\infty$ value and the stage
of cooling (see text) in parentheses; here $t_\infty$ is the time from
the start of the accretion outburst as seen by an observer at
infinity. In addition, for each curve the location and temperature of
the ocean-crust boundary is marked with an open circle. The parameters
used for the analytic approximation are $\tau_b^{(2)} = 13.4$~days,
$\nabla^{(1)} = -0.093$, and $\nabla_{\rm eff}^{(2)} = 0.025$; such
that stage~2 begins (to an outside observer) at $t_{2,\infty} =
17.6$~days and stage~3 begins at $t_{3,\infty} = 63.4$~days. We use
$T_b$ and $y_b$ from the simulations, rather than from
Eqs.~(\ref{eq:Tbanly})--(\ref{eq:ybanly2}), to obtain a better
fit. (Cf.\ figure~2 of Paper~II; note the typographical error in the
``50~days'' curve of that plot.)}
\label{fig:Tcool}
\end{figure}

Figure~\ref{fig:Tcool} shows the temperature profiles at various times
during cooling, along with the analytic approximation to these
profiles. As can be seen in the figure, the ocean-crust boundary moves
outward more quickly during cooling than during accretion:
Equation~(\ref{eq:dotyb}) as it applies to the cooling case is given
by
\be
\dot{y}_b = - \frac{1}{\chi_1/\chi_T + b_1/c_P} \frac{X_{1,b}}{\Delta X_{1,bc}} \frac{F_{r,\rm cd}(y_b^-)-F_{r,\rm cd}(y_b^+)}{c_PT_b} \,;
\label{eq:dotyb2}
\ee
using Eq.~(\ref{eq:Fcd}) with a temperature gradient at the ocean base
$\nabla_b \sim 0.25$ (see below), we find $\dot{y}_b \sim -10^5~{\rm
g~cm^{-2}~s^{-1}}$. This is markedly different from the situation in
Section~\ref{sec:accretion}, where $|\dot{y}_b| \ll \dot{m}$ over most
of the evolution. The composition also evolves more quickly during
cooling than during accretion: From Eq.~(\ref{eq:Xbguess}) we have
$\partial X_{1,b}/\partial t \simeq -\dot{y}_b\Delta X_{1,bc}/y_b$,
which is a factor of $\sim \Delta X_{1,bc}/\Delta X_{1,0c} \agt 20$
times larger than the accretion value [cf.\ Eq.~(\ref{eq:tss})].

\begin{figure}
\begin{center}
\includegraphics[width=\columnwidth]{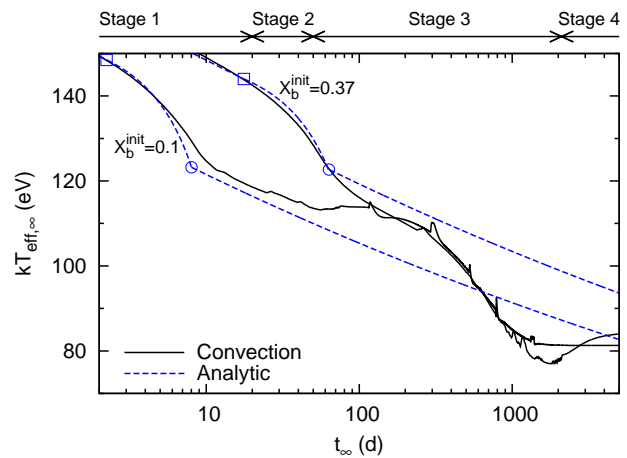}
\end{center}
\caption[The cooling light curve]
{The cooling light curve of a neutron star with $X_b^{\rm init}=0.37$
and compositionally driven convection, and the analytic approximation
to this light curve [Eqs.~(\ref{eq:Teff1}) and (\ref{eq:Teff2})]. Here
$t_\infty$ is the time from the end of the accretion outburst and
$T_{\rm eff,\infty}$ is the effective temperature as seen by an
observer at infinity. The labels that appear above the graph denote
the duration of the stages of cooling (see text). For the analytic
approximation, the transition from stage~1 to 2 and the transition
from stage~2 to 3 are marked with an open square and an open circle,
respectively. The parameters used are the same as in
Fig.~\ref{fig:Tcool}, with the addition of $y_b^{(2)} =
6.05\times10^{13}~{\rm g~cm^{-2}}$ and $T_b^{(2)} =
2.73\times10^8$~K. The light curve and analytic approximation for the
case with $X_b^{\rm init}=0.1$ are also plotted for
comparison. (Cf.\ figure~1 of Paper~II.)}
\label{fig:lightcurve}
\end{figure}

Figure~\ref{fig:lightcurve} shows the cooling light curve along with
the analytic approximation. As can be seen in the figure, changing
$X_{1,b}^{\rm init}$ has a strong effect on the light curve. This is
for two reasons: First, for a larger light-element fraction in the
ocean the thermal conductivity $K \propto \langle Z \rangle^{-1}$ is
also larger; a larger $K$ reduces the temperature gradient in the
ocean (while self-consistently increasing the flux there), which keeps
the outer layers hotter both during steady-state accretion and at the
end of cooling when the crust and core are equilibrated (see
BC09). Second, for a larger light-element fraction in the ocean the
ocean-crust boundary is deeper, which delays the onset of strong ocean
cooling due to compositionally driven convection. Note that in
Fig.~\ref{fig:lightcurve} the analytic approximation deviates strongly
from the model light curve for $t_\infty \agt 100$~days. This is
because our analytic expressions [Eqs.~(\ref{eq:Teff1}) and
(\ref{eq:Teff2})] only account for cooling of the ocean and crust by
heat conduction out through the envelope, not for late-time cooling by
heat conduction into the core (cf.\ BC09).

%%%%
\section{Comparison to observations}
\label{sec:observation}

In Paper~II we presented fits to observations of XTE~J1701--462 and
IGR~J17480--2446, using our model of compositionally driven
convection; here we present fits to observations of several additional
quiescent, transiently accreting neutron stars. Our goal in making
these fits was to understand qualitatively how including convection in
the ocean changes the fitting parameters for these sources. Therefore,
we did not attempt to accurately fit our model to the observational
data using rigorous parameter searches. Similar to BC09, each source
was fit by running our simulations from the onset of accretion through
the duration of the accretion outburst, then turning off accretion and
tracking the cooling light curve out to the end of the observation. In
our fits we take $\dot{m}$ and the duration of the accretion outburst
from observations and fit to the parameters $T_c$, $Q_{\rm imp}$,
$y_0$, and $X_{1,b}^{\rm init}$. Note that in \cite{degenaar14} we
assumed shallow heating in our fit of EXO~0748--676 (see also
BC09). In this paper we do not include shallow heating in our model
directly. Instead, we vary $y_0$ to provide the necessary shallow
heating, with a larger $y_0$ placing the burning layer closer to the
bulk of the ocean and heating it more.

\begin{figure}
\begin{center}
\includegraphics[width=1.05\columnwidth]{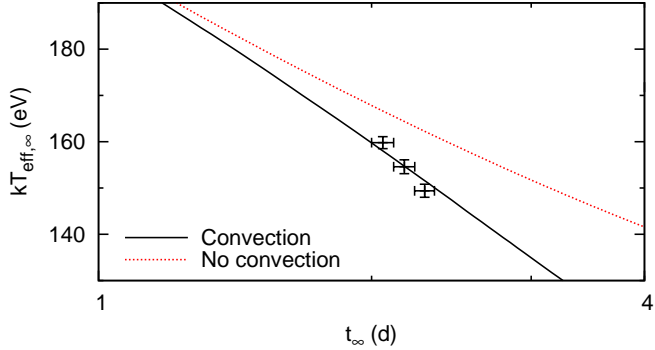}
\end{center}
\vspace{-0.6cm}
\caption[Fitting observations of XTE~J1709--267]
{Model light curves with compositionally driven convection (solid
curve) and without (dotted curve), plotted over the observations of
XTE~J1709--267. The solid curve is a fit to the observations, while
the dotted curve has the same parameters as the solid curve. In our
fit we assume that the first observation was taken 2~days after
accretion turned off. We use $y_0 = 10^{10}~{\rm g~cm^{-2}}$, $T_c =
10^7$~K, $Q_{\rm imp} = 0$, and $X_{1,b}^{\rm init} = 0.03$.}
\label{fig:1709}
\end{figure}

\begin{figure}
\begin{center}
\includegraphics[width=1.05\columnwidth]{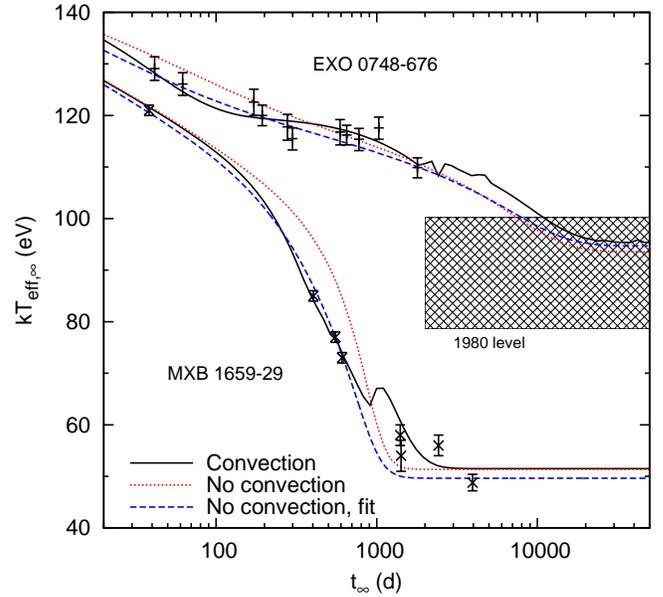}
\end{center}
\vspace{-0.6cm}
\caption[Fitting observations of EXO~0748--676 and MXB~1659--29]
{Model light curves with compositionally driven convection (solid
curves) and without (dashed and dotted curves), plotted over the
observations of EXO~0748--676 and MXB~1659--29. For each source, the
solid curve and the dashed curve are fits to the observations, while
the dotted curve has the same parameters as the solid curve. For
EXO~0748--676, the pre-outburst data ($kT_{\rm eff,\infty} =
94.6_{-16.0}^{+5.6}$~eV) is marked with a shaded bar. For
EXO~0748--676 we set $y_0 = 6\times10^9~{\rm g~cm^{-2}}$ and $Q_{\rm
imp} = 40$, and use $T_c = 1.15\times10^8$~K and $X_{1,b}^{\rm init} =
0.4$ (solid and dotted curves) and $T_c = 1.2\times10^8$~K and
$X_{1,b}^{\rm init} = 0.3$ (dashed curve)
\citep[cf.][]{degenaar14}. For MXB~1659--29 we set $y_0 = 10^9~{\rm
g~cm^{-2}}$ and $T_c = 3\times10^7$~K, and use $Q_{\rm imp} = 4$ and
$X_{1,b}^{\rm init} = 0.8$ (solid and dotted curves) and $Q_{\rm imp}
= 5$ and $X_{1,b}^{\rm init} = 0.55$ (dashed curve). The late-time
wiggles in the EXO~0748--676 convection curve (from $t_\infty \simeq
2000$ to $9000$~days post-outburst) are numerical artifacts; however,
the bump in the MXB~1659--29 convection curve (at $\simeq 1000$~days)
is physical, caused by light-element saturation in the ocean as
described in the text.}
\label{fig:0748y1659}
\end{figure}

Figures~\ref{fig:1709} and \ref{fig:0748y1659} (see also figure~4 of
Paper~II) show our fits to cooling light curves from several quiescent
sources. Note that in most of our fits we use $y_0 \sim 10$ times
larger than the standard value of ${\rm a~few}\times10^8~{\rm
g~cm^{-2}}$ (e.g., \citealt{bildsten97}; Paper~I); i.e., we must
invoke a significant shallow heat source. Convection does not directly
reduce the required shallow heat source for each fit: For the same
shallow heating (same values of $y_0$) the fits for our model with and
without convection are generally equally valid (e.g., in
Fig.~\ref{fig:0748y1659}); in addition, the fitted ocean temperature
at the start of cooling is similar in both our model and that of BC09,
implying the use of a comparable shallow heating model. Instead,
convection justifies the use of larger values of $X_{1,b}$ in our
models due to light-element enrichment, which in turn increases the
thermal conductivity in the ocean and makes it hotter without the need
for shallow heating (see, e.g., the effect of different $X_{1,b}^{\rm
init}$ values in Fig.~\ref{fig:lightcurve}).

In our fits here and in Paper~II, several trends appear when comparing
the light curve from the model with compositionally driven convection
to that from the model without convection, for the same parameters
(see also Fig.~\ref{fig:lightcurve}). First, at $t_\infty \sim
1$--100~days post-outburst the light curve with convection drops below
the light curve without convection then flattens out, as the cooling
transitions from stage~1 to stage~2 to stage~3
(Section~\ref{sec:cooling}). This arises because the
compositionally-driven convection transports heat inwards, rapidly
cooling the ocean and temporarily slowing the cooling in the crustal
layers where the phase separation occurs. Second, at late times the
light curve with convection crosses above the light curve without
convection, due to light-element enrichment during convection
increasing the ocean thermal conductivity (stage~4); for several of
our fits (IGR~J17480--2446 from Paper~II, EXO~0748--676, and
MXB~1659--29) this happens within the observation. Third, our fits
that have steep (shallow) light curves with convection will have
correspondingly steep (shallow) light curves without convection;
compare, e.g., our fits for MXB~1659--29 versus those for
EXO~0748--676 in Fig.~\ref{fig:0748y1659}. This is because, whether or
not compositionally driven convection is in effect, the crust
ultimately drives the cooling (see Section~\ref{sec:cooling}). We do
not discuss the behavior of the crust cooling in this paper; a
detailed discussion can be found in BC09.

\begin{figure}
\begin{center}
\includegraphics[width=1.05\columnwidth]{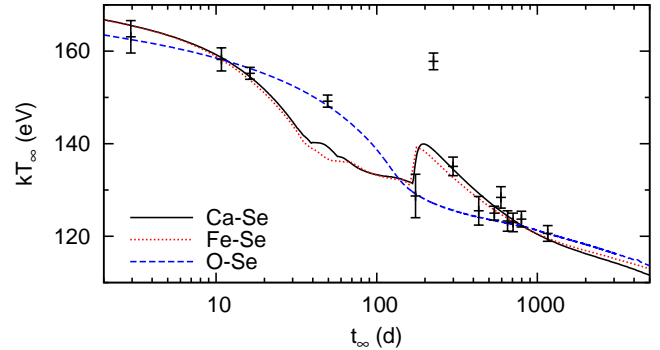}
\end{center}
\vspace{-0.6cm}
\caption[Fitting the spike in XTE~J1701--462]
{Model light curves with compositionally driven convection, plotted
over the observations of XTE~J1701--462. Here we deviate from the
model of Sections~\ref{sec:convection} and \ref{sec:model} by using an
ocean mixture of Ca-Se (solid line) and Fe-Se (dotted line), in place
of O-Se (dashed line). For the Ca-Se ocean we use $y_0 = 10^8~{\rm g~cm^{-2}}$, $T_c
= 1.7\times10^8$~K, $Q_{\rm imp} = 200$, and $X_{1,b}^{\rm init} =
0.5$; for the Fe-Se ocean we use $y_0 = 10^8~{\rm g~cm^{-2}}$, $T_c =
2\times10^8$~K, $Q_{\rm imp} = 200$, and $X_{1,b}^{\rm init} =
0.74$; and for the O-Se ocean we use the parameters from Paper~II,
$y_0 = 5\times10^7~{\rm g~cm^{-2}}$, $T_c = 1.8\times10^8$~K, $Q_{\rm
imp} = 40$, and $X_{1,b}^{\rm init} = 0.37$. The bumps in the Ca-Se
and Fe-Se curves are caused by light-element saturation in the ocean.}
\label{fig:1701Cmp}
\end{figure}

As can be seen from Figs.~\ref{fig:1709} and \ref{fig:0748y1659}, we
generally fit currently available light curves equally well with and
without compositionally driven convection. Detecting the signature of
convection will require better sampling of the early phase of the
cooling curve. XTE~J1709--267 \citep[$\dot{m} = 2\times10^4~{\rm
g~cm^{-2}~s^{-1}}$ with a 10~week outburst;][]{degenaar13b} is the
exception to the above generalization, since the model with convection
fits better to the observed rapid decrease in the cooling light curve
(Fig.~\ref{fig:1709}; cf.\ stage~2 of
Fig.~\ref{fig:lightcurve}). Convection also provides an explanation
for the observed increase in the equilibrium flux level in
IGR~J17480--2446 from 2009 to 2014 (\citealt{degenaar13a}; see
Paper~II), because it allows the composition, and therefore the
equilibrium temperature profile, to change from one accretion episode
to the next. For XTE~J1701--462 (\citealt{fridriksson11}; see
Paper~II), we find that with convection we can match the drop in the
light curve at $100$--$200$~days (Fig.~\ref{fig:1701Cmp}). Similarly
for EXO~0748--676 \citep[$\dot{m} = 2\times10^3~{\rm
g~cm^{-2}~s^{-1}}$ with a 24~year
outburst;][]{degenaar11b,degenaar14}, we find that the inclusion of
convection leads to a plateau of slow cooling between $\simeq
150$--750~days post-outburst, broadly consistent with the data
(Fig.~\ref{fig:0748y1659}). We note that the model with convection is
not statistically preferred over the model without convection.

BC09 fit the light curve of MXB~1659--29 \citep[$\dot{m} =
9\times10^3~{\rm g~cm^{-2}~s^{-1}}$ with a 2.5~year
outburst;][]{wijnands03,wijnands04,cackett08,cackett13} with a
standard cooling model. Using similar parameters and including
convection gives a model light curve with a ``stage~2'' drop at $\sim
50$~days. Because of the gap in the data at $\simeq 40$--400~days, we
have the freedom in our fits to choose where this drop occurs; we can
instead move the drop to $\sim 200$~days by increasing $X_{1,b}^{\rm
init}$ to an unphysical 0.8 (but see below). Note that in our model
there is an abrupt jump in the light curve at late time ($t_\infty
\simeq 1000$~days in Fig.~\ref{fig:0748y1659}), where the base of the
ocean is saturated with light elements and compositionally driven
convection halts (Section~\ref{sec:cooling}). This is a general
feature of our fits to MXB~1659--29, as long as $X_{1,b}^{\rm init}
\agt 0.3$, and it arises due to the steep drop in the light curve
which causes rapid and prolonged outward motion of the ocean-crust
boundary and strong chemical separation ($X_{1,b} \rightarrow 1$). The
observations at late times neither support nor dispute the existence
of this predicted bump (see Fig.~\ref{fig:0748y1659}).

With the model of Sections~\ref{sec:convection} and \ref{sec:model} it
is impossible to fit both ``anomalous'' data points in the light curve
of XTE~J1701--462 (the two points from $t_\infty \simeq 200$ to
$300$~days post-outburst in Fig.~\ref{fig:1701Cmp}; see also figure~4
of Paper~II). However, we can partially fit this data by considering
ocean mixtures other than oxygen-selenium and/or large values of
$X_{1,b}^{\rm init}$. Two such fits, one for a calcium-selenium ocean
and one for an iron-selenium ocean \citep[cf.][]{horowitz07}, are
shown in Fig.~\ref{fig:1701Cmp}; here the light-element saturation in
the ocean produces a bump in the light curve that matches the second
anomalous data point and that has a peak occurring at the same time as
the first anomalous data point. We believe these modifications to our
basic model to be reasonable, considering our uncertainty regarding
what two-component mixture to use for the ocean or whether a
two-component mixture is an accurate representation of the ocean. The
large changes produced in the light curves when different compositions
are used (figure~4 of Paper~II vs. Fig.~\ref{fig:1701Cmp}) emphasizes
the need for models with three or more components. In addition, as we
discuss in Section~\ref{sec:discuss}, it may be possible to reproduce
the amplitude of the rebrightening in XTE~J1701--462 by including
heating due to electron captures self-consistently in our convection
model.

%%%%
\section{Discussion}
\label{sec:discuss}

In this paper we have continued the exploration begun in Paper~I of
the consequences of chemical separation and subsequent compositionally
driven convection in the ocean of accreting neutron stars; while the
model described in Paper~I included only a steady-state ocean, here we
use a full envelope-ocean-crust model and track its behavior from the
onset of accretion to the end of cooling.

We have discovered a strong effect due to compositionally driven
convection on the light curves of cooling, transiently accreting
neutron stars. As the neutron star cools after an accretion outburst,
the ocean-crust boundary moves outward. We find that this leads to
chemical separation, and then convective mixing and inward heat
transport, in a manner similar to that during accretion but at a much
faster rate. The inward heat transport cools the outer layers of the
ocean rapidly, but keeps the inner layers hot; the result is a sharp
drop in surface emission at around a week (depending on parameters),
followed by a gradual recovery as the ocean base moves outward. Such a
dip should be observable in the light curves of these neutron star
transients, if enough data is taken at a few days to a month after the
end of accretion. If such a dip is definitively observed, it will
provide strong constraints on the chemical composition of the ocean
and outer crust.

Enrichment of the ocean with carbon remains a major issue for
superburst models \citep{schatz03}. Following \citet{horowitz07} and
Paper~I, we chose oxygen as the light element for our examples in this
paper, but we have calculated models with carbon as the light element
with similar results (as expected, due to the comparable
heavy-element-to-light-element charge ratios and mass numbers of the
C-Se and O-Se systems which yield comparable phase diagrams and
thermodynamic quantities). We find that chemical separation can enrich
the ocean to the required carbon fraction $X_{\rm C,ign} \simeq 0.1$
\citep{schatz03} within a few months of cooling after an accretion
outburst (cf.\ Section~\ref{sec:cooling}). This is well within the
estimated superburst recurrence time of 1--3~years
\citep{kuulkers02,intzand03}, and is far more efficient than either
chemical separation during accretion heating or gravitational
sedimentation during quiescence. The rapid enrichment during cooling
may help explain the puzzling superburst observed immediately before
the onset of an accretion outburst in EXO~1745--248
\citep{altamirano12}: since the carbon in the ocean is at the required
ignition level before accretion even starts, if the ocean can be
heated strongly enough with a small amount of accretion a superburst
can occur right at the beginning of an outburst.

In order for chemical separation to occur, however, the composition at
the base of the ocean and at the top of the crust must differ; this
will not happen if accretion outbursts are too short to push accreted
material to the base of the ocean. Ultimately the carbon excess is
being supplied by the ashes of the hydrogen and helium burning layer,
with carbon mass fraction $X_{\rm C,0} \alt 0.01$ during unstable
burning \citep{woosley04}. This excess is driven to the ignition depth
$y_{\rm ign} \sim 10^{12}~{\rm g~cm^{-2}}$ within a few months to a
year ($y_{\rm ign}/\dot{m}$); but it takes ten times longer to build
the excess up to the required fraction 0.1
(cf.\ Fig.~\ref{fig:Xevolve}). The total build-up time is at best a
factor of three longer than the estimated superburst recurrence time
of 1--3~years \citep{kuulkers02,intzand03}. We suggest that while a
recurrence rate of a few years can not be sustained through
compositionally driven convection, it is possible to have several
bursts in a row at that rate if a small fraction of carbon can be
``stored'' in the deep ocean or crust (perhaps in lamellar sheets; see
Section~\ref{sec:model}) after each burst. On the other hand,
\citet{intzand03} inferred observationally that stable burning is
happening in superburst sources. Although the physical mechanism for
the stable burning is not understood, it could produce much larger
carbon fractions $X_{\rm C,0} \sim 0.2$ \citep{stevens14}, which would
reduce the timescale needed to enrich the ocean even during accretion
(Section~\ref{sec:accretion}).

Note that while compositionally driven convection may help superburst
models reach the levels of carbon enrichment required for carbon
ignition, it does not help the models reach the required large ocean
temperatures $T_{\rm ign} \sim 6\times10^8$~K \citep{cumming06}. In
fact, we find (Sections~\ref{sec:accretion} and \ref{sec:cooling})
that temperatures in the bulk of the ocean are slightly lower with
convection than without.

Two issues presented in Paper~I have been resolved in the Appendix of
this paper. In Appendix~\ref{sec:mixing} we discuss what happens when
$\nabla > \nabla_{\rm ad}$ in the ocean (see also
Section~\ref{sec:model}). As we alluded to in Paper~I, the small
amount of hydrogen and helium in the transition region between the
burning layer and the ocean stabilize the density gradient at the top
of the ocean and allows for an unstable temperature gradient and
heavy-element composition gradient simultaneously, such that there is
no contradiction between a small convective velocity at the top of the
ocean and a smooth composition transition from the burning layer to
the ocean. In Appendix~\ref{sec:rotmag} we discuss how rotation and
magnetic fields affect our model. We find that the efficient
convection assumption Eq.~(\ref{eq:nablaXe}) remains valid even in the
presence of rapid rotation and moderate magnetic fields; the remaining
temperature and composition evolution equations in the paper follow
directly from it and are therefore also unaffected by rotation or
magnetic field.

There remains much to be explored theoretically. We have included only
two species in our calculations, oxygen and selenium, which
approximates the rp-process ashes used by \cite{horowitz07}. The phase
diagram for multicomponent mixtures is complex but can be calculated
\citep{horowitz07,medin10} and should be included. Multiple,
consecutive, accretion outburst-quiescence cycles should also be
simulated to obtain self-consistent composition profiles in the ocean
and outer crust. It will be important to include carbon burning in the
models.

We have assumed that solid particles form at a single depth. However,
electron capture reactions may occur in the ocean (e.g., ${}^{56}$Fe
captures at a density of $1.5\times 10^9~{\rm g~cm^{-3}}$;
\citealt{haensel90}), lowering the $\langle Z \rangle$ at that depth,
and potentially leading to formation of solid particles pre-electron
capture above the post-electron capture liquid layers. The heat
released due to electron captures during mixing and sedimentation of
the region could be observable in the light curve. Further work is
needed to understand how electron captures would affect the model
presented here.

\acknowledgements

We thank Chuck Horowitz, Nathalie Degenaar, and Chris Fontes for
useful discussions. Z.M. was supported by a LANL Director's
Postdoctoral Fellowship. A.C. is supported by an NSERC Discovery
grant, and is a member of the Centre de Recherche en Astrophysique du
Qu\'ebec (CRAQ) and an Associate of the CIFAR Cosmology and Gravity
program. This research was carried out in part under the auspices of
the National Nuclear Security Administration of the U.S. Department of
Energy at Los Alamos National Laboratory and supported by Contract
No. DE-AC52-06NA25396.

\appendix

%%%%
\section{Convective stability}
\label{sec:stable}

Here we derive expressions for the convective discriminant ${\cal A}$
and other quantities related to entropy production and convective
stability in the multicomponent oceans of neutron stars.

The usual stability requirement for a displaced fluid element is
\citep[e.g.,][]{kippenhahn94}
\be
{\cal A} < 0 \,,
\label{eq:stab}
\ee
where
\be
{\cal A} = \frac{d\ln \rho}{dr} - \left(\frac{d\ln \rho}{dr}\right)_{s,X_i,Y_e}
\label{eq:Aconv}
\ee
with $d/dr$ the gradient in the star and $(d/dr)_{s,X_i,Y_e}$ the
gradient felt by an element displaced at constant entropy $s$ and
chemical composition $\{X_i,Y_e\}$ (i.e., we assume the element is
displaced with no radiated energy and no chemical diffusion). In the
neutron star ocean, the sound speed is much larger than the convective
velocity, such that a displaced element is always in pressure
balance with its surroundings:
\be
\frac{d\ln P}{dr} = \left(\frac{d\ln P}{dr}\right)_{s,X_i,Y_e} \,.
\label{eq:dlnPdr}
\ee
Using Eq.~(\ref{eq:dlnPdr}) and 
\be
d\ln P = \chi_T d\ln T + \chi_\rho d\ln \rho + \sum_{i=1}^n \chi_{X_i} d\ln X_i + \chi_{Y_e} d\ln {Y_e} \,,
\label{eq:dlnP}
\ee
we can rewrite ${\cal A}$ as
\be
{\cal A} = -\frac{1}{\chi_\rho}\left[\chi_T \frac{d\ln T}{dr} + \sum_{i=1}^n \chi_{X_i} \frac{d\ln X_i}{dr} + \chi_{Y_e}\frac{d\ln Y_e}{dr} -\chi_T \left(\frac{d\ln T}{dr}\right)_{s,X}\right] \,;
\ee
defining
\be
\chi_i = \chi_{X_i}-\chi_{X_n}\frac{X_i}{X_n}+\chi_{Y_e}\frac{(Y_i-Y_n)X_i}{Y_e}
\ee
and enforcing the constraints $\sum_{i=1}^n X_i = 1$ and $Y_e =
\sum_{i=1}^n Y_iX_i$, the convective discriminant becomes
\be
{\cal A} = \frac{1}{H_P\chi_\rho}\left[\chi_T (\nabla-\nabla_{\rm ad}) + \sum_{i=1}^{n-1} \chi_i \nabla_{X_i}\right] \,.
\label{eq:Aconv2}
\ee

Using
\be
\left(\frac{\partial s}{\partial T}\right)_{P,X_i,Y_e} = \frac{c_P}{T} \,,
\ee
\be
\left(\frac{\partial s}{\partial P}\right)_{T,X_i,Y_e} = -\left(\frac{\partial T}{\partial P}\right)_{s,X_i,Y_e}\left(\frac{\partial s}{\partial T}\right)_{P,X_i,Y_e} = -\frac{\nabla_{\rm ad}}{P}c_P \,,
\ee
and Eqs.~(\ref{eq:bPi}) and (\ref{eq:bPe}), we have
\be
ds = c_P d\ln T - c_P\nabla_{\rm ad} d\ln P - \sum_{i=1}^n b_{P,i} d\ln X_i - b_{P,e} d\ln {Y_e} \,;
\label{eq:ds}
\ee
defining
\be
b_i = b_{P,i}-b_{P,n}\frac{X_i}{X_n}+b_{P,e}\frac{(Y_i-Y_n)X_i}{Y_e}
\ee
and again enforcing $\sum_{i=1}^n X_i = 1$ and $Y_e = \sum_{i=1}^n
Y_iX_i$, we can rewrite Eq.~(\ref{eq:ds}) as
\be
ds = c_P d\ln T - c_P\nabla_{\rm ad} d\ln P - \sum_{i=1}^{n-1} b_i d\ln X_i \,.
\label{eq:ds2}
\ee
Therefore
\be
\frac{ds}{dr} = -\frac{1}{H_P} \left[c_P (\nabla-\nabla_{\rm ad}) - \sum_{i=1}^{n-1} b_i \nabla_{X_i}\right] \,.
\label{eq:dsdr}
\ee
Assuming that the pressure at a given depth does not change with time
\citep[cf.\ appendix~A of][]{brown98} we also have from
Eq.~(\ref{eq:ds2}) that
\be
\frac{\partial s}{\partial t} = \frac{c_P}{T}\frac{\partial T}{\partial t} - \sum_{i=1}^{n-1} \frac{b_i}{X_i}\frac{\partial X_i}{\partial t} \,.
\label{eq:dsdt}
\ee

%%%%
\section{Mixing length equations and efficient convection}
\label{sec:mixing}

Here we derive or define expressions for several quantities related to
heat transfer and composition mixing, first using mixing length theory
and then using the efficient convection assumption of
Eq.~(\ref{eq:nablaXe}). We discuss the the regimes in which either
model is appropriate. Finally we discuss how to make these models
consistent with an ocean that has $\nabla > \nabla_{\rm ad}$.

In mixing length theory \citep[e.g.,][]{kippenhahn94}, a displaced
element feels an average force per unit mass of
\be
-\frac{gD\rho}{2\rho} = -\frac{g}{2\rho} \left[\left(\frac{d\rho}{dr}\right)_{s,X_i,Y_e} - \frac{d\rho}{dr}\right] = \tfrac{1}{2}g{\cal A}l_m
\label{eq:fconv}
\ee
applied over an average distance of $l_m/2$; assuming that
approximately half of this work goes into the kinetic energy of the
particle, the convective velocity is given by
\be
v_{\rm conv}^2 = c_s^2\frac{\xi^2}{8\chi_\rho}\left(\chi_T(\nabla - \nabla_{\rm ad}) + \sum_{i=1}^{n-1} \chi_i\nabla_{X_i}\right) = \tfrac{1}{8}g{\cal A} l_m^2
\label{eq:vconv}
\ee
where $c_s = (gH_P)^{1/2}$ is the sound speed and $\xi = l_m/H_P$ is
the ratio between the convection mixing length $l_m$ and the scale
height (but see Appendix~\ref{sec:rotmag}). The composition flux for
species $i$ is given by
\be
F_{r,X_i} \hat{\mathbf{r}} = \rho v_{\rm conv} DX_i \hat{\mathbf{r}}
\label{eq:Fx}
\ee
where the composition ``excess'' of the displaced element over its
surroundings is
\be
DX_i = -\frac{l_m}{2}\frac{dX_i}{dr} = \frac{\xi}{2}X_i\nabla_{X_i} \,;
\ee
the convective heat flux is given by
\be
F_{r,\rm conv} \hat{\mathbf{r}} = \rho v_{\rm conv} TDs \hat{\mathbf{r}}
\label{eq:F}
\ee
where
\be
Ds = -\frac{l_m}{2}\frac{ds}{dr} = \frac{\xi}{2}c_P\left(\nabla-\nabla_{\rm ad} - \frac{1}{c_P}\sum_{i=1}^{n-1} b_i \nabla_{X_i}\right)
\label{eq:sexcess}
\ee
using Eq.~(\ref{eq:dsdr}). To solve for the evolution of the ocean
using mixing length theory, we assume a value for $\xi$ and use
Eqs.~(\ref{eq:vconv})--(\ref{eq:sexcess}) to find $F_{r,X_i}$ and
$F_{r,\rm conv}$ for Eqs.~(\ref{eq:xflux}) and (\ref{eq:EBE}).

For efficient convection [Eq.~(\ref{eq:nablaXe})], Eq.~(\ref{eq:dsdr})
becomes
\be
T\dot{m}\frac{\partial s}{\partial y} = -\frac{c_PT\dot{m}}{y} \sum_{i=1}^{n-1} \frac{\chi_i}{\chi_T}\left(1 + \frac{\chi_Tb_i}{\chi_ic_P}\right) \nabla_{X_i} \,,
\label{eq:eTmdsdy}
\ee
while Eq.~(\ref{eq:F}) with Eq.~(\ref{eq:Fx}) becomes
\be
F_{r,\rm conv} = -\frac{c_PT}{\chi_T}\sum_{i=1}^{n-1} \frac{\chi_i}{\chi_T}\left(1 + \frac{\chi_Tb_i}{\chi_ic_P}\right) \frac{F_{r,X_i}}{X_i} \,.
\label{eq:AeF}
\ee
Using Eqs.~(\ref{eq:xflux}), (\ref{eq:eTmdsdy}), and (\ref{eq:AeF})
with $\epsilon_X = 0$, we have
\be
\frac{\partial F_{r,\rm conv}}{\partial y} = -\sum_{i=1}^{n-1} \left\{\frac{c_PT\chi_i}{X_i\chi_T}\left(1 + \frac{\chi_Tb_i}{\chi_ic_P}\right) \left[\frac{\partial X_i}{\partial t} + \dot{m}\frac{\partial X_i}{\partial y}\right] + F_{r,X_i} \frac{\partial}{\partial y} \left[\frac{c_PT\chi_i}{X_i\chi_T}\left(1 + \frac{\chi_Tb_i}{\chi_ic_P}\right)\right] \right\} \,,
\label{eq:dFcvdy}
\ee
such that with Eqs.~(\ref{eq:Fr}), (\ref{eq:dsdt}), and
(\ref{eq:eTmdsdy}) the energy balance equation Eq.~(\ref{eq:EBE})
becomes
\be
c_P\frac{\partial T}{\partial t} + \sum_{i=1}^{n-1} \frac{c_PT\chi_i}{X_i\chi_T}\frac{\partial X_i}{\partial t} = \frac{\partial F_{r,\rm cd}}{\partial y} -\sum_{i=1}^{n-1} F_{r,X_i} \frac{\partial}{\partial y} \left[\frac{c_PT\chi_i}{X_i\chi_T}\left(1 + \frac{\chi_Tb_i}{\chi_ic_P}\right)\right] + \epsilon \,.
\label{eq:AEBE2}
\ee
To solve for the evolution of the ocean using the efficient convection
assumption we use the procedure described in
Section~\ref{sec:model}. Note that
Eqs.~(\ref{eq:eTmdsdy})--(\ref{eq:AEBE2}) are independent of $\xi$,
such that we do not need to assume a value for this parameter.

\begin{figure}
\begin{center}
\begin{tabular}{cc}
\includegraphics[width=0.5\columnwidth]{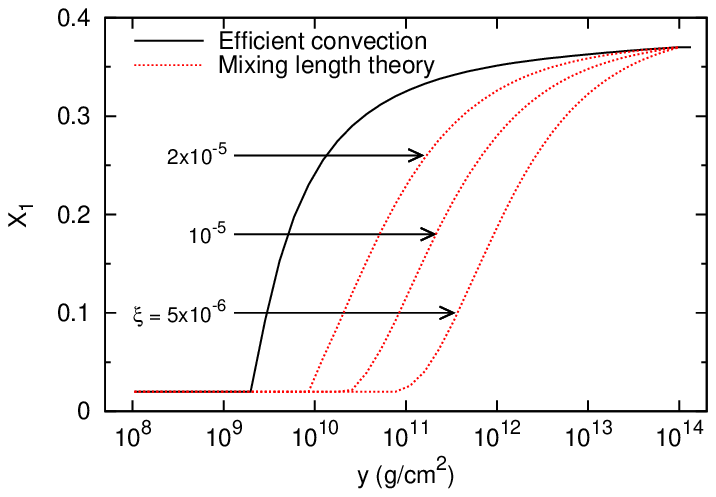} &
\includegraphics[width=0.5\columnwidth]{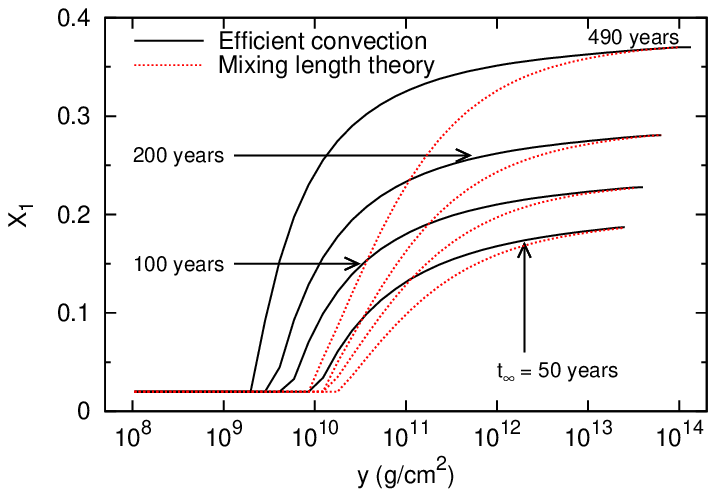}
\end{tabular}
\end{center}
\caption[The effect of $\xi$ on the composition profile]
{Composition profiles in the ocean of a neutron star with
compositionally driven convection, in mixing length theory for various
values of $\xi$ and in the efficient convection assumption. The left
panel shows the steady-state composition profiles as $\xi$ increases
from $5\times10^{-6}$ to $2\times10^{-5}$; the right panel shows the
composition profile at various times for $\xi=2\times10^{-5}$ and in
the efficient convection assumption.}
\label{fig:xicomp}
\end{figure}

Figure~\ref{fig:xicomp} shows the composition profile for the example
from Section~\ref{sec:accretion}, using mixing length theory with
various values of $\xi$, and using the efficient convection
assumption. The value of $\xi$ at which efficient convection becomes a
good approximation in the neutron star ocean can be estimated using
Eq.~(\ref{eq:xfluxb}) as an upper bound for the composition flux:
Eq.~(\ref{eq:Fx}) gives
\be
\frac{\xi}{2}\rho v_{\rm conv,max} \simeq \frac{\dot{m}-\dot{y}_b}{\sum_{i=1}^{n-1} \chi_i\nabla_{X_i}} \sum_{i=1}^{n-1}\chi_i\frac{\Delta X_{i,bc}}{X_i} \sim 10^7~{\rm g~cm^{-2}~s^{-1}} \,;
\ee
using Eq.~(\ref{eq:vconv}), $v_{\rm conv} \le v_{\rm conv,max}$, and
$\dot{m}-\dot{y}_b \sim 10^5~{\rm g~cm^{-2}~s^{-1}}$ we have
\be
\frac{\sum_{i=1}^{n-1} \chi_i\nabla_{X_i}}{\chi_T\left(\nabla_{\rm ad}-\nabla\right)} - 1 \le \frac{32\chi_\rho}{\xi^4} \left(\frac{\frac{\xi}{2}\rho v_{\rm conv,max}}{\rho c_s}\right)^2 \sim \left(\frac{10^{-5}}{\xi}\right)^4 \,.
\ee
This means that for mixing length parameters $\xi \gg 10^{-5}$ (as we
assumed in Paper~I), convection is efficient; i.e., $\sum_{i=1}^{n-1}
\chi_i\nabla_{X_i}$ is extremely close to its maximum stable value,
$\chi_T\left(\nabla_{\rm ad}-\nabla\right)$. Note that when $\xi \gg
10^{-5}$, small numerical errors in $\chi_T\left(\nabla - \nabla_{\rm
ad}\right) + \sum_{i=1}^{n-1} \chi_i\nabla_{X_i}$ lead to very large
errors in $v_{\rm conv}$. In this case we can not use the full mixing
length procedure but must assume efficient convection.

In Paper~I we suggested that a time-dependent calculation could help
resolve what happens when a stable composition profile can not extend
from the burning layer ash at the top of the ocean to the steady-state
mixture at the ocean base (i.e., from $X_{1,0} = 0.02$ to $X_{1,b} =
0.37$ for the ${}^{16}$O-${}^{79}$Se system); this could happen, e.g.,
when $\nabla > \nabla_{\rm ad}$ at the top of the ocean. In the
current paper we remove the inconsistency by allowing the composition
at the top of the ocean to be different from that provided by the
burning layer (e.g., Fig.~\ref{fig:Xburning}; see also figure~3 from
Paper~II), and use the stabilizing effect of the burning layer on
convection as justification. If we instead fix
$\left\{X_{1,0}\right\}$, we find that once the convection zone
reaches the top of the ocean there is a flux at the outer boundary
\be
F_{r,X_i}(y_0) = \dot{m} \Delta X_{i,0c}
\label{eq:FX02}
\ee
[Eqs.~(\ref{eq:xflux}) and (\ref{eq:xfluxb})], with $X_{i,0} \ne
X_{i,c}$ because the system is not yet in steady state. In the O-Se
system, this means that a large quantity of oxygen is being ejected
from the ocean into the envelope, a fact that we are ignoring because
of our assumption of a fixed envelope composition. We conclude that
the only way to make our models consistent with a $\nabla >
\nabla_{\rm ad}$ ocean is to consider the envelope, by either allowing
ocean material to mix into the envelope [through Eq.~(\ref{eq:FX02})],
or by including hydrogen and helium burning in the envelope to prevent
mixing (as in Fig.~\ref{fig:Xburning}).

%%%%
\section{Heat and composition sources}
\label{sec:sources}

Here we derive expressions for the sources $\epsilon_{X_i}$ and
$\epsilon$, as used in the continuity equation Eq.~(\ref{eq:xflux})
and entropy balance equation Eq.~(\ref{eq:EBE}), respectively.

There are sources of composition change $\epsilon_{X_i}$ at three
locations in our model:

1) At the ocean-crust boundary, composition changes abruptly due to
chemical separation and rapid sedimentation of the solid at the phase
transition. Formally, we write
\be
\epsilon_{X_i} = -(\dot{m}-\dot{y}_b) \Delta X_{i,bc} \delta(y-y_b) \,,
\ee
where $\delta$ is the Dirac delta function; but in practice we simply
adjust $X_{i,c}$ manually without reference to Eq.~(\ref{eq:xflux}).

2) At the hydrogen and helium burning layer, composition changes
quickly due to the strong temperature dependence of the thermonuclear
reactions, from $\{X_{i,e}\}$ in the envelope to $\{X_{i,0}\}$ at the
top of the ocean. We assume that the burning layer is infinitely thin,
such that formally we have
\be
\epsilon_{X_i} = \dot{m} (X_{i,0}-X_{i,e}) \delta(y-y_0) \,.
\ee
Note that $X_{i,0}$ is not necessarily the value given by the burning
layer ashes ($\{X_{\rm O},X_{\rm Se}\} = \{0.02,0.98\}$ for the O-Se
ocean model of this paper); we allow $X_{i,0}$ to vary based on the
composition profile required by efficient convection in the ocean (see
Section~\ref{sec:model}).

3) In the crust, composition changes gradually due to electron
captures and pycnonuclear fusion. For simplicity we set $\langle A
\rangle_c = 56$ as the composition at the top of the crust, regardless
of the value of $X_{i,c}$, and follow the procedure of BC09 to obtain
$\langle Z \rangle$ and $\langle A \rangle$ at greater depths. With
this approximation, $X_{i,c}$ only determines the physics of the
liquid-solid phase transition [e.g., in Eq.~(\ref{eq:Xbguess})] and
has no effect on the crust properties (thermal conductivity, etc.).

During accretion, there are heat sources $\epsilon$ at three locations
in our model (see Fig.~\ref{fig:Fprofile}):

1) At the hydrogen and helium burning layer, particles are driven to a
critical depth (temperature) for thermonuclear reactions by accretion;
for $\{X_{\rm H},X_{\rm He}\} = \{0.7,0.3\}$ these reactions release
$Q = 5~{\rm MeV~nucleon^{-1}}$ \citep[e.g.,][]{brown98}. Following
BC09, we assume that the heat is released uniformly in the logarithm
of column depth, over a region from $y = y_{\rm low}$ to $y_{\rm
high}$, such that
\be
\epsilon = \left\{
\begin{array}{ll}
\dfrac{Q\dot{m}/m_p}{y \ln (y_{\rm high}/y_{\rm low})} \,, & y_{\rm low} < y < y_{\rm high} \,; \\
0 \,, & \mbox{otherwise.}
\end{array}
\right.
\label{eq:epsilon}
\ee
Here we choose $y_{\rm low} = 0.2y_0$ and $y_{\rm high} = y_0$.

2) In the outer crust, electron captures release $Q = 0.2~{\rm
MeV~nucleon^{-1}}$ \citep[e.g.,][]{haensel08}; we use $y_{\rm low} =
5\times10^{12}~{\rm g~cm^{-2}}$ and $y_{\rm high} =
5\times10^{15}~{\rm g~cm^{-2}}$.

2) In the inner crust, pycnonuclear fusion reactions release $Q =
1.2~{\rm MeV~nucleon^{-1}}$; we use $y_{\rm low} = 5\times10^{15}~{\rm
g~cm^{-2}}$, and $y_{\rm high} = 3\times10^{18}~{\rm g~cm^{-2}}$.

%%%%
\section{Thermodynamic quantities}
\label{sec:thermo}

Here we derive or define expressions for several thermodynamic
quantities in multicomponent plasmas that are used in this paper.

The total differential for the Gibbs free energy is given by
\be
dG = -SdT + VdP + \sum_{i=1}^n \mu_i dN_i + \mu_e dN_e \,,
\label{eq:Gtdiff}
\ee
where $G$ includes the energy of the ions and the electrons, $V$ is
the total volume, $n$ is the total number of ion species in the
plasma, $\mu_i$ is the chemical potential of ion species $i$, $N_i$ is
the number of ions of species $i$, $\mu_e$ is the electron chemical
potential, and $N_e$ is the number of electrons. Note that although
$N_e = \sum_{i=1}^n Z_i N_i$ for the fully-ionized multicomponent
plasma, such that $N_e$ is not an independent thermodynamic variable,
here we treat it as such in order to express the various relations
derived in this section in terms of both ion and electron
quantities. The ion and electron terms are combined in the rest of the
paper [Eqs.~(\ref{eq:chi1}) and (\ref{eq:b1})] to simplify the
appearance of the equations. Using $X_i = A_i N_i/\langle A \rangle
N$, $Y_e = \sum_{i=1}^n Y_iX_i$, and the Euler integral
\be
G = \sum_{i=1}^n \mu_iN_i + \mu_eN_e \,,
\ee
where $G$ is the Gibbs free energy, we have
\be
dg = -sdT + \frac{1}{\rho}dP + \sum_{i=1}^n \frac{\mu_i}{A_im_p}dX_i + \frac{\mu_e}{m_p}dY_e \,.
\label{eq:gtdiff}
\ee
Here $q$ is the ``specific'' version of the quantity $Q$; i.e., $q =
Q/M$, where $M = \langle A \rangle m_p N$ is the total mass of the
system and $N = \sum_{i=1}^n N_i$ is the total number of ions. Using
Eq.~(\ref{eq:gtdiff}) we can derive two useful Maxwell relations:
Since
\be
\left(\frac{\partial^2 g}{\partial X_i \partial T}\right)_{P,X_{j \ne i},Y_e} = \left(\frac{\partial^2 g}{\partial T \partial X_i}\right)_{P,X_{j \ne i},Y_e} \,,
\ee
we have
\be
-\left(\frac{\partial s}{\partial X_i}\right)_{T,P,X_{j \ne i},Y_e} = \frac{1}{A_im_p}\left(\frac{\partial \mu_i}{\partial T}\right)_{P,X_i,Y_e} \,;
\label{eq:maxi}
\ee
similarly,
\be
-\left(\frac{\partial s}{\partial Y_e}\right)_{T,P,X_i} = \frac{1}{m_p}\left(\frac{\partial \mu_e}{\partial T}\right)_{P,X_i,Y_e} \,.
\label{eq:maxe}
\ee

We define
\be
b_{P,i} \equiv -X_i \left(\frac{\partial s}{\partial X_i}\right)_{T,P,X_{j \ne i},Y_e} = \frac{X_i}{A_im_p}\left(\frac{\partial \mu_i}{\partial T}\right)_{P,X_i,Y_e}
\label{eq:bPi}
\ee
and
\be
b_{P,e} \equiv -Y_e \left(\frac{\partial s}{\partial Y_e}\right)_{T,P,X_i} = \frac{Y_e}{m_p}\left(\frac{\partial \mu_e}{\partial T}\right)_{P,X_i,Y_e} \,;
\label{eq:bPe}
\ee
these terms are analogous to the ion and electron specific heat terms
$c_{P,i} = T(\partial s_i/\partial T)_{T,P,X_i}$ and $c_{P,e} =
T(\partial s_e/\partial T)_{T,P,Y_e}$ for the composition. For
degenerate electrons
\be
b_{P,e} = -\frac{\pi^2k_BT}{3E_F}\frac{Y_ek_B}{m_p} = -\tfrac{2}{3}c_{P,e} \,,
\label{eq:bPeest1}
\ee
where $E_F = m_ec^2\left(\sqrt{1+x_F^2}-1\right)$ is the Fermi energy
with $x_F = 10.0 \rho_9^{1/3} Y_e^{1/3}$; for the models we consider
here the electrons are degenerate and Eq.~(\ref{eq:bPeest1}) holds
throughout the ocean, since $k_BT/E_F \alt 0.2$ (for $\rho \agt
10^6~{\rm g~cm^{-3}}$ and $T \simeq 3\times10^8$~K). An accurate
expression for $b_{P,i}$ in the ocean can be obtained from $\mu_i =
(\partial F_l/\partial N_i)_{T,V,N_{j \ne i}}$ and the free energy of
a multicomponent liquid
\be
F_l = k_BT \sum_{i=1}^n N_i \left[f_l^{\rm OCP}(\Gamma_i) + \ln\left(\frac{N_iZ_i}{N_e}\right)\right] \,,
\ee
where $f_l^{\rm OCP}$ (including the ideal gas part) is defined in
equations~1 and 2 of \cite{medin10}; we use this accurate expression
in our numerical calculations. However, an approximation for $b_{P,i}$
can be obtained by considering only the ideal gas term, the dominant
temperature-dependent term in $\mu_i$:
\be
b_{P,i} \simeq \frac{X_i}{A_im_p} k_B\ln\left[\frac{N_i}{V}\left(\frac{h^2}{2\pi A_im_pk_BT}\right)^{3/2}\right] \sim 30 \frac{X_ik_B}{A_im_p} \sim 10 c_{P,i} \,.
\label{eq:bPiest}
\ee

%%%%
\section{Tracking the ocean-crust boundary}
\label{sec:tracking}

Here we derive expressions for the motion of the ocean-crust boundary,
as well as the changes in entropy and composition at the ocean crust
boundary.

Using Eq.~(\ref{eq:yb}) and
\be
\langle Z_b^{5/3} \rangle = \sum_{i=1}^n x_{i,b} Z_i^{5/3} = \langle A \rangle \sum_{i=1}^n X_{i,b}\frac{Z_i^{5/3}}{A_i} \,,
\ee
we have that the ocean-crust boundary moves at a rate
\be
\dot{y}_b = \dot{y}_{b,T} + \dot{y}_{b,X} \,,
\label{eq:Adotyb3}
\ee
where
\be
\dot{y}_{b,T} = \frac{\partial y_b}{\partial T_b}\frac{\partial T_b}{\partial t} = \frac{4y_b}{T_b}\frac{\partial T_b}{\partial t}
\label{eq:dotybT}
\ee
and
\bal
\dot{y}_{b,X} = {}& \sum_{i=1}^n \frac{\partial y_b}{\partial X_{i,b}}\frac{\partial X_{i,b}}{\partial t} = \sum_{i=1}^{n-1} y'_{b,i}\frac{\partial X_{i,b}}{\partial t}
\label{eq:dotybX}
\eal
with
\be
\frac{\partial y_b}{\partial X_{i,b}} = \frac{4y_b \langle A \rangle}{\langle Z_b^{5/3} \rangle}\frac{\langle Z_b^{5/3} \rangle - Z_i^{5/3}}{A_i}
\ee
and
\be
y'_{b,i} = \frac{\partial y_b}{\partial X_{i,b}}-\frac{\partial y_b}{\partial X_{n,b}} \,.
\ee
For an ${}^{16}$O-${}^{79}$Se mixture, $y'_{b,1} \simeq 10y_b$. At the
ocean-crust boundary, the energy balance equation Eq.~(\ref{eq:EBE})
in the frame moving with the boundary is
\be
T_b\frac{\partial s_b}{\partial t} + T_b(\dot{m}-\dot{y}_b)\left.\frac{\partial s}{\partial y}\right|_{y=y_b^{\rm down}} = \left.\frac{\partial F_{r}}{\partial y}\right|_{y=y_b} + \epsilon \,,
\label{eq:EBEb}
\ee
where $y_b^{\rm down}$ indicates that the derivative is evaluated on
the ``downstream'' side of the boundary: $y_b^+$ during heating
($\dot{y}_b > 0$), and $y_b^-$ during cooling ($\dot{y}_b < 0$). Note
that $(\partial s/\partial y)_{y=y_b^+} = 0$ and that $\epsilon = 0$
during cooling.

For a two component mixture, near the base of the ocean $\nabla_{X_1}
= \chi_T\left(\nabla_{\rm ad}-\nabla\right)/\chi_1 \ll 1$
(cf.\ Paper~I), such that $\partial X_1/\partial t$ is almost constant
there. Therefore, using Eq.~(\ref{eq:Xinout}),
\be
\frac{\partial X_{1,b}}{\partial t} \simeq \frac{\dot{m}\Delta X_{1,0c} - \dot{y}_b\Delta X_{1,bc}}{y_b} \,;
\ee
with Eqs.~(\ref{eq:Adotyb3})--(\ref{eq:dotybX}) we have
\be
\frac{\partial X_{1,b}}{\partial t} \simeq \frac{1}{1+y'_{b,1}\Delta X_{1,bc}/y_b} \left(\frac{\dot{m}\Delta X_{1,0c}}{y_b} - \frac{4\Delta X_{1,bc}}{T_b} \frac{\partial T_b}{\partial t}\right) \,.
\label{eq:AdotX}
\ee
During cooling we have from Eqs.~(\ref{eq:Xinout}) and
(\ref{eq:AdotX}) that
\be
4\frac{\partial \ln T_b}{\partial t} \simeq \left(1+\frac{y'_{b,1}\Delta X_{1,bc}}{y_b}\right) \frac{\partial \ln y_b}{\partial t} \,;
\label{eq:AdotT}
\ee
for $\Delta X_{1,bc} \ll 1$, $4(\partial \ln T_b/\partial t) \simeq
\partial \ln y_b/\partial t$; while for $\Delta X_{1,bc} \sim 1$,
$\partial \ln T_b/\partial t \sim 3(\partial \ln y_b/\partial t$).

%%%%
\section{Effects of rotation and magnetic field on convection}
\label{sec:rotmag}

The effects of rotation and magnetic fields on convection have been
examined in many places (e.g.,
\citealt{stevenson79,stevenson03,jones00,christensen06}; see also
\citealt*{showman11}). Here we use simple arguments to show that in
the neutron star ocean, the efficient convection assumption
Eq.~(\ref{eq:nablaXe}) is very good even in the presence of rapid
rotation ($\sim 10^2~{\rm s^{-1}}$) and moderate magnetic fields
($\sim 10^{10}$~G).

We consider a two-component ocean mixture with a plane-parallel
geometry and governed by Newtonian physics. We impose a gravitational
field $-g\hat{\mathbf{r}}$, rotation $\mathbf{\Omega}$, and magnetic
field $\mathbf{B}_0$, all uniform. We assume that during convective
mixing, displaced fluid elements do not exchange heat or material with
their surroundings until they have traveled a distance of order the
mixing length $l_m$; but by rapidly contracting or expanding they
maintain pressure balance with their surroundings
(cf.\ Appendix~\ref{sec:stable}). We therefore have
\be
\left(\frac{dP_{\rm tot}}{dr}\right)_{s,X,Y_e} = \frac{dP_{\rm tot}}{dr} \,,
\ee
where $P_{\rm tot} = P + P_{\rm mag}$ and
\be
P_{\rm mag} = \frac{B^2}{8\pi} \,.
\ee
Note that the equations of
Appendices~\ref{sec:thermo}--\ref{sec:mixing} are unaffected by the
inclusion of this magnetic ``pressure'' term because $P_{\rm mag} \ll
P$. In the rotating frame, the equation of motion for a fluid element
displaced from its equilibrium position $\mathbf{r}_0$ is
\be
\rho \frac{\partial \mathbf{v}}{\partial t} = f_{\rm grav} + f_{\rm rot} + f_{\rm mag}
\ee
where
\be
\mathbf{v} = \frac{\partial \delta\mathbf{r}}{\partial t} \,,
\ee
\be
f_{\rm grav} \simeq -g \left[\left(\frac{d\rho}{dr}\right)_{s,X,Y_e} - \frac{d\rho}{dr}\right] \delta r_r \hat{\mathbf{r}} \simeq \rho g {\cal A} \delta r \hat{\mathbf{r}}
\ee
is the buoyancy force (per unit volume) felt by the element,
\be
f_{\rm rot} = -2\rho \mathbf{\Omega}\times\mathbf{v}
\ee
is the Coriolis force, and
\be
f_{\rm mag} = \frac{1}{4\pi} (\nabla\times\mathbf{B})\times\mathbf{B} + \nabla P_{\rm mag} = \frac{1}{4\pi} (\mathbf{B}\cdot\nabla)\mathbf{B} \simeq \frac{1}{4\pi} (\mathbf{B}_0\cdot\nabla)\delta\mathbf{B}
\ee
is the magnetic ``tension'' force. Here
\be
\delta\mathbf{r} = \mathbf{r} - \mathbf{r}_0
\ee
is the displacement of the element from its equilibrium position and
\be
\delta\mathbf{B} = \mathbf{B} - \mathbf{B}_0
\ee
is the perturbation to the magnetic field caused by this
displacement. The magnetic tension term $f_{\rm mag}$ acts as a
restoring force \citep[cf.][]{kulsrud05}: for a typical wavelength
$l_m$ and a perpendicular field displacement of $\delta\mathbf{r}$ we
have that
\be
\delta\mathbf{B} \sim \frac{B_0}{l_m} \delta\mathbf{r}
\ee
such that
\be
f_{\rm mag} \sim \rho \omega_B^2 \delta\mathbf{r} \,,
\ee
where
\be
\omega_B^2 = \frac{B_0^2}{4\pi \rho l_m^2} = \left(\frac{v_A}{l_m}\right)^2
\ee
and
\be
v_A = \frac{B_0}{\sqrt{4\pi \rho}}
\ee
is the Alfv\'en velocity.

We assume that the rotation and magnetic fields are oriented in no
particular direction relative to each other or the gravitational
field: $\mathbf{\Omega} = \Omega_x \hat{\mathbf{x}} + \Omega_r
\hat{\mathbf{r}}$ and $\mathbf{B}_0 = B_{0,x} \hat{\mathbf{x}} +
B_{0,y} \hat{\mathbf{y}} + B_{0,r} \hat{\mathbf{r}}$, where $\Omega_x
\sim \Omega_r$ and $B_{0,x} \sim B_{0,y} \sim B_{0,r}$. We then have
the equations of motion
\be
\frac{\partial^2 \delta r}{\partial t^2} \sim (g {\cal A} - \omega_B^2) \delta r + 2\Omega \frac{\partial \delta x}{\partial t}
\label{eq:eomr}
\ee
and
\be
\frac{\partial^2 \delta x}{\partial t^2} \sim -2\Omega \frac{\partial \delta r}{\partial t} - \omega_B^2 \delta x
\label{eq:eomx}
\ee
(with the assumptions made above, $\delta y$ is independent of $\delta
r$ and $\delta x$, and so we do not consider it further). If we assume
that both $\delta r$ and $\delta x$ depend on $t$ as $\exp(\sigma t)$,
where $\sigma$ is a constant, Eqs.~(\ref{eq:eomr}) and (\ref{eq:eomx})
give
\be
\sigma^4 + (2\omega_B^2 + 4\Omega^2 - g {\cal A})\sigma^2 \sim \omega_B^2 (g {\cal A} - \omega_B^2) \,.
\label{eq:sigma}
\ee

Equations~(\ref{eq:Fx}) and (\ref{eq:F}) for the composition and
convective heat fluxes still apply when $B_0 \ne 0$ and $\Omega \ne
0$, except that $v_{\rm conv}$ there is now the convective velocity
only in the radial direction. Equation~(\ref{eq:vconv}) no longer
applies, however, since the force per unit mass on the displaced
element is not given by just the buoyancy term
Eq.~(\ref{eq:fconv}). Instead we use
\be
v_{{\rm conv},r} = \left.\frac{\partial \delta r}{\partial t}\right|_{\delta r = l_m/2} = \tfrac{1}{2}\sigma l_m \,.
\label{eq:vconvgen}
\ee
The convective velocity must be large enough to carry the required
composition flux Eq.~(\ref{eq:xfluxb}); $\sigma$, the oscillation
frequency for the convective instability, will grow until this
happens. Comparing Eqs.~(\ref{eq:Fx}) and (\ref{eq:xfluxb}), we have
that
\be
\sigma l_m^2 \simeq \frac{4(\dot{m}-\dot{y}_b)}{\rho} \frac{\Delta X_{bc} H_P}{X \nabla_X} \,,
\ee
or assuming $l_m \sim H_P$ and $\dot{m}-\dot{y}_b \sim 10^5~{\rm
g~cm^{-2}~s^{-1}}$ (cf.\ Appendix~\ref{sec:mixing}),
\be
v_{{\rm conv},r} \sim 10^{-2}~{\rm cm~s^{-1}} \qquad{\rm and}\qquad \sigma \sim 10^{-5}~{\rm s^{-1}} \,.
\ee
For transiently accreting neutron stars $\Omega \sim 10^2~{\rm
s^{-1}}$ and $B_0 \alt 10^{10}$~G, such that $\Omega \agt \omega_B \gg
\sigma$. Therefore we have from Eq.~(\ref{eq:sigma}) that
\be
g{\cal A} \simeq \omega_B^2 \,,
\label{eq:Agen}
\ee
and that
\be
\sigma^2 \sim \frac{\omega_B^2}{\omega_B^2+4\Omega^2} (g {\cal A} - \omega_B^2) \,.
\label{eq:sigma2}
\ee
From Eq.~(\ref{eq:sigma2}) we see that $\sigma < 0$, and therefore
convection is inhibited, until the convective discriminant is at least
as large as $\omega_B^2/g$; i.e., until the buoyancy force exceeds the
magnetic tension force. A slight excess of $g{\cal A}$ over
$\omega_B^2$ gives the composition flux necessary to transport the
chemical imbalance at the base of the ocean.

Since ${\cal A} = \chi_T\left(\nabla-\nabla_{\rm ad}\right)/(\chi_\rho
H_P) + \chi_X\nabla_X/(\chi_\rho H_P) \simeq \omega_B^2/g \sim
10^{-10}~{\rm cm^{-1}}$ while $\chi_T\left(\nabla_{\rm
ad}-\nabla\right)/(\chi_\rho H_P) \sim 10^{-6}~{\rm cm^{-1}} \gg {\cal
A}$, we have
\be
\chi_X\nabla_X \simeq \chi_T\left(\nabla_{\rm ad}-\nabla\right) \,;
\ee
i.e., the efficient convection assumption Eq.~(\ref{eq:nablaXe}) is
good even in the presence of rotation and magnetic fields.

Note that there is some ambiguity in the typical length scale for the
problem. Putting Eq.~(\ref{eq:sigma2}) back into Eqs.~(\ref{eq:eomr})
and (\ref{eq:eomx}) gives
\be
\delta x \sim \frac{2\Omega \sigma}{\omega_B^2} \delta r \,;
\ee
since $\sigma \ll \omega_B$, the typical perturbation in the
horizontal direction is much smaller than in the vertical
direction. This may require an average displacement smaller than
$l_m/2$ to be used in Eq.~(\ref{eq:vconvgen}) \citep[see,
e.g.,][]{stevenson79}, which will increase the oscillation frequency
$\sigma$ required to generate the composition flux $F_X$. However, we
still have $\sigma^2 \ll g{\cal A}$ such that $g{\cal A}$ will not
change much (unless $\omega_B$ also changes) and our conclusion
remains the same. Note also that if the star is non-magnetic such that
$\omega_B = 0$, Eq.~(\ref{eq:sigma}) instead gives
\be
\sigma^2 \sim g {\cal A} - 4\Omega^2
\ee
and we have
\be
{\cal A} \simeq 4\Omega^2/g \sim 10^{-10}~{\rm cm^{-1}} \,;
\ee
we again find that efficient convection is a good assumption.

\end{document}